%% file: main.tex
\documentclass[12pt]{article}

\usepackage{mathpazo}
\usepackage[colorlinks, linkcolor=blue, citecolor=blue]{hyperref}           
\usepackage{color,mathrsfs}
\usepackage{cite,graphicx,subfigure,amsmath,amssymb,amsfonts,bm,epsfig,epsf,url,dsfont}
\usepackage{amsthm}
\usepackage{tikz}
\usepackage{bbm}      
\usepackage{booktabs}
\usepackage{cases}
\usepackage{fullpage}
\usepackage[small,bf]{caption}
\usepackage[top=1in,bottom=1in,left=1in,right=1in]{geometry}
\usepackage{fancybox}

\usepackage{enumerate}
\usepackage[shortlabels]{enumitem}

\usepackage{algorithm}
\usepackage{verbatim}
\usepackage{algorithmic}

\usepackage{mathtools}

\usepackage{scalerel,stackengine}
\stackMath
\newcommand\reallywidehat[1]{%
\savestack{\tmpbox}{\stretchto{%
  \scaleto{%
    \scalerel*[\widthof{\ensuremath{#1}}]{\kern-.6pt\bigwedge\kern-.6pt}%
    {\rule[-\textheight/2]{1ex}{\textheight}}
  }{\textheight}%
}{0.5ex}}%
\stackon[1pt]{#1}{\tmpbox}%
}

\input{macros}
\makeatletter
\newcommand*{\rom}[1]{\expandafter\@slowromancap\romannumeral #1@}
\makeatother

\newtheoremstyle{exampstyle}
  {0.1cm} 
  {0cm} 
  {\itshape} 
  {} 
  {\bfseries} 
  {.} 
  {.5em} 
  {} 

\theoremstyle{exampstyle}\newtheorem{question}{Question}

\usepackage{enumerate}

\newcommand{\Ind}{\mathbbm{1}}

\newcommand{\s}[1]{\mathsf{#1}}

\newcommand{\abs}[1]{\left|#1\right|}

\input{myshorts}

\newcommand{\p}[1]{\left(#1\right)}
\newcommand{\pp}[1]{\left[#1\right]}
\newcommand{\ppp}[1]{\left\{#1\right\}}
\newcommand{\norm}[1]{\left\|#1\right\|}

\newcommand{\indep}{\perp \!\!\! \perp}

\allowdisplaybreaks

\usepackage{fancybox}

\begin{document}

\title{Inferring Hidden Structures in Random Graphs}

\author{Wasim~Huleihel\thanks{W. Huleihel is with the Department of Electrical Engineering-Systems at Tel Aviv university, {T}el {A}viv 6997801, Israel (e-mail:  \texttt{wasimh@tauex.tau.ac.il}). This research was supported by the ISRAEL SCIENCE FOUNDATION (grant No. 1734/21).}
}

\maketitle

\begin{abstract}

We study the two inference problems of detecting and recovering an isolated community of \emph{general} structure planted in a random graph. The detection problem is formalized as a hypothesis testing problem, where under the null hypothesis, the graph is a realization of an Erd\H{o}s-R\'{e}nyi random graph $\calG(n,q)$ with edge density $q\in(0,1)$; under the alternative, there is an unknown structure $\Gamma_k$ on $k$ nodes, planted in $\calG(n,q)$, such that it appears as an \emph{induced subgraph}. In case of a successful detection, we are concerned with the task of recovering the corresponding structure. For these problems, we investigate the fundamental limits from both the statistical and computational perspectives. Specifically, we derive lower bounds for detecting/recovering the structure $\Gamma_k$ in terms of the parameters $(n,k,q)$, as well as certain properties of $\Gamma_k$, and exhibit computationally unbounded optimal algorithms that achieve these lower bounds. We also consider the problem of testing in polynomial-time. As is customary in many similar structured high-dimensional problems, our model undergoes an ``easy-hard-impossible" phase transition and computational constraints can severely penalize the statistical performance. To provide an evidence for this phenomenon, we show that the class of low-degree polynomials algorithms match the statistical performance of the polynomial-time algorithms we develop. 
\end{abstract}

\section{Introduction}

The past decade has seen the emergence of datasets of an unprecedented scale, with both large sample size and dimensionality. Massive datasets arise in various domains, among them are computer vision, natural language processing, computational biology, and social networks analysis, to name a few. Any solution to a machine learning problem has two central aspects: \emph{statistical} and \emph{computational}. The statistical aspect characterizes the performance of desired inference tasks, while the computational aspect studies the computational complexity of efficient algorithms constructed for these tasks. For many years, the investigation of the two aspects has largely happened in isolation, for the sake of modularity.

Traditionally, information theory and statistics have been the main framework to understand the statistical aspect. Since the influential work of \cite{LC86,IK81,yang1999,Aad2000}, etc., it has long been recognized that information-theoretic quantities such as entropy and mutual information, as well as bounding methodologies based on Fano's inequality, play an important role in establishing the minimax rates of estimation. A crucial aspect absent from classical statistical analysis is the issue of computational complexity. This aspect, however, is becoming increasingly relevant because, while the sample size and dimensionality of modern datasets seem to grow without bounds, computation is struggling to keep up. 

Over the last few years, there has been a success in developing a rigorous notion of what can and cannot be achieved by efficient algorithms. Recent results, e.g., \cite{berthet2013complexity,ma2015computational,cai2015computational,krauthgamer2015semidefinite,hajek2015computational,chen2016statistical,wang2016average,wang2016statistical,gao2017sparse,brennan18a,brennan19,wu2018statistical,brennan20a,hopkins2017bayesian,Hopkins18,Kunisky19,Cherapanamjeri20,gamarnik2020lowdegree,barak2016nearly,deshpande2015improved,meka2015sum,TengyuWig15,kothari2017sum,hopkins2016integrality,raghavendra2019highdimensional,hopkins2017power,mohanty2019lifting,Feldman17,feldman2018complexity,Diakonikolas17,DiakonikolasKong19,Lenka16,Lesieur_2015,Lesieur_2016,Krzakala10318,Ricci_Tersenghi_2019,bandeira2018notes} revealed an intriguing phenomenon that is common to many high-dimensional problems with a planted structure: there is an inherent gap between the amount of data needed by all computationally efficient algorithms and what is needed for statistically optimal algorithms. Various forms of rigorous evidence for this phenomenon, i.e., hardness in the statistical sense, have been proposed, and they can roughly be classified into two groups: 1) \emph{Failure under certain computation models}, namely, showing that powerful classes of computationally efficient algorithms, such as, low-degree polynomials \cite{hopkins2017bayesian,Hopkins18,Kunisky19,Cherapanamjeri20,gamarnik2020lowdegree}, sum-of-squares hierarchy \cite{barak2016nearly,deshpande2015improved,meka2015sum,TengyuWig15,kothari2017sum,hopkins2016integrality,raghavendra2019highdimensional,hopkins2017power,mohanty2019lifting}, statistical query algorithms \cite{Feldman17,feldman2018complexity,Diakonikolas17,DiakonikolasKong19}, message-passing algorithms \cite{Lenka16,Lesieur_2015,Lesieur_2016,Krzakala10318,Ricci_Tersenghi_2019,bandeira2018notes}, etc., fail in the conjectured computationally hard regime of the problem. 2) \emph{Average-case reductions} from another problem, such as the planted clique problem, conjectured to be computationally hard, e.g., \cite{berthet2013complexity,ma2015computational,cai2015computational,chen2016statistical,hajek2015computational,wang2016average,wang2016statistical,gao2017sparse,brennan18a,brennan19,wu2018statistical,brennan20a}.

Despite the recent progress in understanding the statistical-computational tradeoffs exhibited in several contemporary high-dimensional problems, many fundamental questions remain open. 
Indeed, the ``zoo" of statistical problems with gaps contains a broad range of very different settings and structures. Specifically, most planted structures considered so far typically have a very specific form and are somewhat ad hoc, while there are many important scenarios and applications where these structures can be quite general. Accordingly, existing techniques used to determine the statistical and computational limits are specialized to handle specific types of structures. For example, current reductions have mainly designed for inference problems with structures similar to the starting hardness assumption of planted clique, and as so most techniques are not yet capable of reducing between problems with different high-dimensional structures. The goal of this paper is to develop a comprehensive program aiming to advance the understanding of the fundamental inferential and algorithmic limits of statistical inference on large domains/networks, by investigating detection and recovery of \emph{general} structures planted in random graphs/matrices.

\subsection{Problem Setup}\label{subsec:setup}

As discussed above, most planted structures considered so far typically have a very specific form and are somewhat ad hoc. For example, in the general submatrix detection and recovery problems, e.g., \cite{brennan19}, which subsumes many settings studied in the literature, the planted structure $S$ is a \emph{set} whose elements are \emph{all} either fully connected (clique) or connected by chance; the underlying (expected) graph adjacency matrix of $S$ corresponds to a \emph{complete subgraph}. Below, we put forward a natural way of extending this framework to planted structures modeled by general \emph{graphs}.

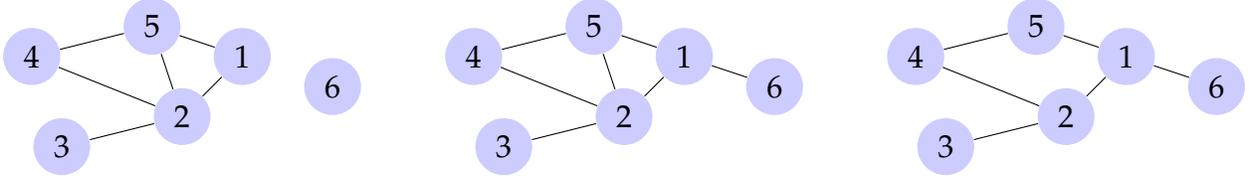
\begin{figure*}[t!]
\centering
\begin{tikzpicture}
  [scale=.4,auto=left,every node/.style={circle,fill=blue!20}]
  \node (n6) at (14,7) {6};
  \node (n4) at (4,8)  {4};
  \node (n5) at (8,9)  {5};
  \node (n1) at (11,8) {1};
  \node (n2) at (9,6)  {2};
  \node (n3) at (5,5)  {3};

  \foreach \from/\to in {n4/n5,n5/n1,n1/n2,n2/n5,n2/n3,n2/n4}
    \draw (\from) -- (\to);

\end{tikzpicture}~~~~~~~~~~~
\begin{tikzpicture}
  [scale=.4,auto=left,every node/.style={circle,fill=blue!20}]
  \node (n6) at (14,7) {6};
  \node (n4) at (4,8)  {4};
  \node (n5) at (8,9)  {5};
  \node (n1) at (11,8) {1};
  \node (n2) at (9,6)  {2};
  \node (n3) at (5,5)  {3};

  \foreach \from/\to in {n4/n5,n5/n1,n1/n2,n2/n5,n2/n3,n2/n4,n6/n1}
    \draw (\from) -- (\to);
\end{tikzpicture}~~~~~~~~~~~
\begin{tikzpicture}
  [scale=.4,auto=left,every node/.style={circle,fill=blue!20}]
  \node (n6) at (14,7) {6};
  \node (n4) at (4,8)  {4};
  \node (n5) at (8,9)  {5};
  \node (n1) at (11,8) {1};
  \node (n2) at (9,6)  {2};
  \node (n3) at (5,5)  {3};

  \foreach \from/\to in {n4/n5,n5/n1,n1/n2,n2/n3,n2/n4,n6/n1}
    \draw (\from) -- (\to);

\end{tikzpicture}
\caption{Base graph (left), union ensemble (middle), subgraph ensemble (right).}
\label{fig:2}
\end{figure*}

Let us describe the setting we plan to study, starting with the detection problem. We have a total population of $n$ individuals. Let $k\in\mathbb{N}$, and fix a graph $\Gamma_k = ([k],\calE_{\Gamma_k})$ with node set $[k]\triangleq\{1,2,\ldots,k\}$ and edge set $\calE_{\Gamma_k}$. Let $\pi:[k]\to[n]$ be an injective map chosen uniformly at random. Given $\Gamma_k$, denote by $\s{H}_{\Gamma_k}=(\calN_{\s{H}_{\Gamma_k}},\calE_{\s{H}_{\Gamma_k}})$ the graph with $n$ nodes $\calN_{\s{H}_{\Gamma_k}}\triangleq\{\pi(i):i\in[k]\}$ and edges $\calE_{\s{H}_{\Gamma_k}}\triangleq\{(\pi(i),\pi(j)):(i,j)\in\calE_{\Gamma_k}\}$. We shall refer to $\s{H}_{\Gamma_k}$ (or, $\Gamma_k$) as the \emph{planted/hidden} structure. Consequently, our detection problem can be phrased as the following simple hypothesis testing problem: under the uniform hypothesis $\calH_0$, the graph $\s{G}$ is an Erd\H{o}s-R\'{e}nyi random graph $\calG(n,q)$ with edge density $0<q<1$, which might be a function of $n$. Under the planted hypothesis $\calH_1$, we sample first a base graph $\s{G}' = ([n],\calE')$ from $\calG(n,q)$, and then we \emph{plant} the graph $\s{H}_{\Gamma_k} = (\calN_{\s{H}_{\Gamma_k}},\calE_{\s{H}_{\Gamma_k}})$ in $\s{G}'$. There are two meaningful possible ways to plant $\s{H}_{\Gamma_k}$. The first, is to take the union of the base graph with $\s{H}_{\Gamma_k}$, i.e., $\s{G}=\s{G}'\cup\s{H}_{\Gamma_k}$; we refer to this as the \emph{union ensemble}. The second ensemble, which we focus on, is constructed as follows: 
\begin{enumerate}
\itemsep-0.3em
    \item We \emph{remove} all the edges between the vertices $\calN_{\s{H}_{\Gamma_k}}$ in the base graph $\s{G}'$.
    \item We add the edges $\calE_{\s{H}_{\Gamma_k}}$ in $\s{G}'$. 
\end{enumerate}
The resultant graph is $\s{G}$. An \emph{equivalent} procedure for this construction is:
\begin{enumerate}[\hspace{\parindent}1'.]
\itemsep-0.3em
    \item Take $\s{H}_{\Gamma_k}$ and connect its vertices $\calN_{\s{H}_{\Gamma_k}}$ to the rest of the vertices $[n]\setminus\calN_{\s{H}_{\Gamma_k}}$ with probability (w.p.) $q$.
    \item Inter-connect the vertices $[n]\setminus\calN_{\s{H}_{\Gamma_k}}$ w.p. $q$.
\end{enumerate}
The procedures above guarantee that the structure $\s{H}_{\Gamma_k}$ appears as an \emph{induced subgraph} under the alternative. We refer to this as the \emph{subgraph ensemble}, and denote the ensemble of random graphs formed by this process by $\calG(n,q,k,\Gamma_k)$. In short, we have the following hypothesis testing problem:
\begin{align}
\calH_0: \s{G} \sim \calG(n,q) \quad \s{vs.} \quad \calH_1 : \s{G} \sim \calG(n,q,k,\Gamma_k).\label{eqn:super_hypo}   
\end{align}
The difference between the two planting procedures is illustrated in Fig.~\ref{fig:2}, where a star configuration is planted on vertices $\{1,2,5,6\}$, with vertex labeled ``$1$" designating the origin. It is evident that in the subgraph ensemble, the star configuration appears as an induced subgraph, while this is not the case for the union ensemble, due to the existence of an edge between vertices ``$2$" and ``$5$". 

We study the above framework in the asymptotic regime where $n\to \infty$, and $k,q$ may also change as a function of $n$. \emph{In this paper, we focus mainly on the already non-trivial dense regime where $q$ is near constant}, i.e., $n^{-o(1)}\leq q\leq 1-n^{-o(1)}$. Observing $\s{G}$, the goal is to design a test/algorithm $\phi(\s{G}) \in \{0, 1\}$ that distinguishes between $\calH_0$ and $\calH_1$. Specifically, the average $\mathsf{Type}$ $\mathsf{I}$+$\mathsf{II}$ risk of a
test $\phi$ is defined as
$\gamma_n (\phi) = \pr_{\calH_0}(\phi(\s{G}) = 1)+ \pr_{\calH_1}(\phi(\s{G}) = 0)$. We say that a sequence of tests $\phi_n$ indexed by $n$ is asymptotically powerful (resp., powerless) if $\gamma_n(\phi_n)\to 0$ (resp., $\to1$). Note that, a sequence of tests is asymptotically powerless if it does not perform any better than random guessing that ignores $\s{G}$. The above is summarized in the following definition.
\begin{definition}[Strong detection]\label{def:detection}
Let $\pr_{\calH_0}$ and $\pr_{\calH_1}$ be the distributions of $\s{G}$ under the uniform and planted hypotheses, respectively. A possibly randomized algorithm $\phi_n(\s{G}) \in \{0, 1\}$ is powerful (i.e., achieves detection) if its $\mathsf{Type}$ $\mathsf{I}$+$\mathsf{II}$ risk satisfies $  \limsup_{n\to\infty}\gamma_n (\phi_n)=0$.
\end{definition}
We remark that the above criterion is known as strong detection, in contrast to weak detection which requires the asymptotic $\mathsf{Type}$ $\mathsf{I}$+$\mathsf{II}$ to be bounded away from unity. In the \emph{recovery} task, upon observing a graph $\s{G}$, drawn from the distribution of $\calG(n,q,k,\Gamma_k)$, one is required to determine a subgraph $\hat{\Gamma}_k(\s{G})$, on $k$ nodes. 
We focus on the following recovery guarantee.
\begin{definition}[Exact recovery]\label{def:recovery}
An algorithm $\hat{\Gamma}_k(\s{G})$ exactly recovers $\Gamma_k$, if, $\pr_{\calH_1}[\hat{\Gamma}_k(\s{G})=\Gamma_k]\to1$, as $n\to\infty$.
\end{definition}

\subsection{Connection to the Literature and Motivation}\label{subsec:litreature}

\textbf{Union vs. subgraph.} The union ensemble described above was studied very recently in \cite{massoulie19a}, for the special case of $\s{D}$-ary trees superimposed in a \emph{sparse} Erd\H{o}s-R\'{e}nyi random graph. One of the main insights/results in \cite{massoulie19a} is that in stark contrast to previously studied statistical problems with a hidden structure, where a rich and intriguing statistical-computation gap emerges, there is no hard phase in the union model. While this phenomenon is surprising it actually makes sense. Consider, for example, the case of planting a star graph. To keep the exposition simple, we focus on the sparse regime where $q=c/n$, for some $c>0$. Nonetheless, the conclusions below remain the same for any $q$. In the union ensemble, it is clear that over $\calH_0$ the expected number of $k$-stars is $n\cdot\binom{n-1}{k}q^{k}$, and this number tends to $\infty$ if $k\ll\log n/\log\log n$. By the first moment method, this gives an intuition as to why for such values of $k$ detecting the planted star is information theoretically impossible. On the other hand, for $k\gg\log n/\log\log n$, an \emph{efficient} test which decides $\calH_1$ if there is a node in $\s{G}$ with degree at least $k$ succeeds with high probability. Recall that in the planted clique problem with $q=1/2$, detection is impossible if $k=o(\log n)$, and statistically hard if $k=o(\sqrt{n})$. For stars, however, there is no such a hard phase (even for $q=1/2$). 

We suspect that the union model in \cite{massoulie19a} originates from the planted clique problem, where there is no need to remove any edge, i.e., the union and subgraph ensembles coincide. While in first glance it might seem that the difference between the two ensembles is semantic only, it turns out that they behave very differently. To see why, consider again the star configuration under the proposed subgraph ensemble with $q=1/2$. Here, since the planted star appears as an induced subgraph, its complement is a clique with an additional disconnected vertex. Therefore, the statistical and computational limits of both are essentially the same! In particular, detecting a star over the subgraph ensemble is as hard as detecting a clique, and accordingly all we know about cliques apply for stars as well. Therefore, in a strike contrast to the union ensemble, where it is easy to detect a star whenever statistically possible, the subgraph ensemble undergoes an ``easy-hard-impossible" phase transition.

\vspace{0.1cm}
\noindent\textbf{Theoretical \& practical motivations.} Our main motivation is theoretical; recent works have shown that statistical inference of planted signals and structures in graphs/matrices undergo an ``easy-hard-impossible" phase transition. In spite of this recent progress of understanding, many fundamental questions remain open. In particular, as mentioned before, the ``zoo" of statistical problems contains a broad range of very different settings and structures, which were not studied hitherto. Our model captures some of these models in the context of random graphs, and we believe that our work elucidates and explains some of the causes for the above type of phase transitions.

Besides the theoretical importance, one possible practical motivation/application of our model is the following \cite{massoulie19a}. Consider the case where $\s{G}'$ models a normal communication system among $n$ individuals. The planted structure, on the other hand, can model a (new) set of anonymous malicious attackers who are connected among themselves in a certain topology to coordinate their activity. Our task is then to detect those set of attackers when they exist, and identify them if there is an attack. This application can in fact be modeled by both the union and subgraph ensembles. The difference is, however, that in the union ensemble the prior communication topology chosen by the malicious users might be destroyed after it is being planted in the graph (e.g., edges which \emph{do not exist} in the chosen topology might be added after taking the union whenever they exist in the base graph). In the subgraph ensemble, on the other hand, the chosen topology is respected by the planting procedure -- edges that exist (not exist) in the topology will exist (not exist) at the end of the planting procedure. We believe that in some applications the subgraph ensemble might be more reasonable at least from a security point of view. For example, the line topology $1\to2\to3$ on three vertices, may represent a hierarchy where agent ``2" is the trustee while agents ``1" and ``3" are kept secret one from another. In fact, an application of the union model for transactional evidence of terrorist activities was already proposed in \cite{Mifflin04}; the background graph represents a large number of entities (vertices) and transactions (edges). The planted subgraph is a simple model for a predetermined pattern of terrorist transactions. Likewise, the background graph is a simple model for massive amounts of ``transactional noise" that make the terrorist pattern difficult to detect.

\vspace{0.1cm}
\noindent\textbf{Related work.} The detection and recovery problems of planted clique have been studied from many different theoretical angles, e.g., \cite{alon1998finding,dekel2014finding,montanari2015finding,barak2016nearly}. A folklore result in the study of these problems is both detection and reconstruction are conjecturelly hard for cliques of size $k=o(\sqrt{n})$ while statistically solvable when $k=\Omega(\log n)$. Information-theoretic thresholds as well as the analysis of efficient algorithms for planted dense and sparse subgraph detection are developed in, e.g., \cite{butucea2013detection,arias2014community,verzelen2015community}. The recovery counterpart of this model was studied in \cite{chen2016statistical}. In the recent years, the planted clique conjecture has been used through reduction arguments to show hardness results in other related high-dimensional problems with planted structure, such as, sparse PCA \cite{berthet2013complexity}, and dense subgraph detection \cite{hajek2015computational}; both of these problems are subject to ``easy-hard-impossible" phases. A systematic treatment of such average-case reductions was initiated in \cite{brennan18a,brennan19,brennan20a}.

The emergence of fixed subgraphs in random graphs has been comprehensively studied in the literature. This dates back to early 60's where Erd\H{o}s and R\'{e}nyi, in their fundamental papers \cite{erdos59a,erdos_renyi_60}, posed the question on the distribution of small subgraphs in $\calG(n,q)$. This problem was subsequently considered in \cite{bollobas_1981,Andrzej85,SPENCER1990286}, and many other papers. The books \cite{bollobas_2001,janson2011random} are devoted to a comprehensive survey of results on the distribution of small subgraphs in $\calG(n,q)$ and to other asymptotic properties of this graph. In a nutshell, Erd\H{o}s and R\'{e}nyi proved in \cite{erdos_renyi_60} that $q = n^{-1/m(\s{H})}$ is the threshold for a copy of $\s{H}$ appearing when $\s{H}$ is balanced, where $m(\s{H})$ denotes the maximum subgraph density (see, eq. \eqref{eqn:maxDensity} for a precise definition), and \cite{bollobas_1981} extended this result to general graphs. Also in \cite{Andrzej85,SPENCER1990286}, the number of copies of $\s{H}$ at the threshold was shown to have Poisson distribution when $\s{H}$ is strictly balanced. Our paper leverage classical lower-bounding techniques based on the second moment method, which were used in above papers, to analyze detection infeasibility which is, in some sense, equivalent to the absence of copies of the planted set in $\calG(n,q)$. Note that, however, the arguments used in the above papers, and, in particular, the statistics of the count of the number of copies of a given structure in $\calG(n,q)$, are more relevant to the union ensemble; while we insist that the structure will appear as an induced subgraph in $\calG(n,q)$, in the above works as well as in the union ensemble the structure may not appear as an induced subgraph. For many planted subgraphs this changes the count distribution dramatically.

Until now, the study of detection and recovery of planted subgraphs over random graphs have been limited to specific structures, such as, cliques, independent sets, and ``random" cliques (i.e., random dense/sparse subgraphs). Most closely related to our work is \cite{massoulie19a}, where the union ensemble was studied for the special case of $\s{D}$-ary trees superimposed in a \emph{sparse} Erd\H{o}s-R\'{e}nyi random graph. Another exception is \cite{Bagaria20}, where the problem of planted Hamiltonian cycle recovery was addressed in \cite{Bagaria20}.

\subsection{Main Contributions}

To the best of our knowledge, the subgraph ensemble introduced in Subsection~\ref{subsec:setup} is novel and has not been studied before, and it opens up many exciting directions for future study of both theoretical and practical significance. In particular, we believe that studying this setting improves our understanding of what causes/drives statistical-computational gaps. Our main contributions in this paper are:
\begin{itemize}[leftmargin=*]

\item\textbf{\underline{Statistical limits.}} We start our investigation of the subgraph ensemble from a statistical point of view, keeping computation considerations aside.
\begin{question}\label{q:1}
Consider the detection and recovery problems in Definitions~\ref{def:detection} and \ref{def:recovery}, respectively. What property of the hidden structure $\Gamma_k$ the statistical limits depend on?
\end{question}
We provide an answer to Question~\ref{q:1}, for general planted structures. We do so by deriving sharp detection and recovery thresholds; on the one hand, we derive an information-theoretic bound that applies to all algorithms, providing conditions under which all algorithms are powerless. On the other hand, we display algorithms that basically achieve the best performance possible. 


\item\textbf{\underline{Inferring in polynomial-time.}} The algorithms achieving the statistical limits are based on an exhaustive (or, combinatorial) search over the solution space and are thus computationally intractable. Accordingly, the next step is to understand what can be achieved in a reasonable time or computational complexity. This is captured by the following question.
\begin{question}\label{q:2}
Construct and analyze poly-time algorithms solving the detection and recovery problems in Definitions~\ref{def:detection} and \ref{def:recovery}, respectively.
\end{question}
We propose two efficient polynomial-time algorithms; the first is based on a simple global degree-count, while the second is a spectral algorithm. We analyze the performance of both algorithms, and observe a substantial gap between the performance of these algorithms and the optimal one.

\item\textbf{\underline{Statistical-computational gaps.}} As mentioned above, we observe a gap between the statistical limits we derive and the performance of the efficient algorithms we construct. We conjecture that this gap is in fact inherent, namely, below the computational barrier polynomial-time algorithms do not exist. To provide an evidence for this conjecture we follow a recent line of work \cite{hopkins2017bayesian,Hopkins18,Kunisky19,Cherapanamjeri20,gamarnik2020lowdegree} and show that the class of low-degree polynomials fail to solve the detection problem in this conjectureally hard regime.
\begin{question}\label{q:3}
What property of $\Gamma_k$ makes it easy (or, hard) to attain the these limits? What can/cannot be achieved by, for example, the class of low-degree polynomials algorithms (see, e.g., \cite{Hopkins18})?
\end{question}
We show that in the regime where $q$ is near constant, if we take degree-$\log n$ polynomials as a proxy for $n^{O(\log n)}$-time algorithms, then $n^{O(\log n)}$-time algorithm do not exist when $k=O(n^{1/2-\epsilon})$, for any $\epsilon>0$.
\end{itemize}

\subsection{Notation and Paper Organization}
In this paper, we adopt the following notational conventions. We denote the size of any finite set $\calS$ by $|\calS|$. For $n\in\mathbb{N}$ we let $[n] = \{1,\ldots,n\}$, and $\binom{[n]}{k}$ denote the set of all size $k$ subsets of $[n]$. For a subset $\calS\subseteq\mathbb{R}$, let $\Ind\pp{\calS}$ denote the indicator function of the set $\calS$. We denote by $\s{Bern}(p)$ and $\s{Binom}(n,p)$ the Bernoulli and binomial distributions with $n$ trials and success probability $p$, respectively. For a simple undirected graph $\s{G} = (\calV,\calE)$, let $v(\s{G}) = |\calV|$ and $e(\s{G}) = |\calE|$ denote the number of vertices and edges in $\s{G}$, respectively. The automorphism group of a graph $\s{G}$, is denoted by $\s{Aut}(\s{G})$. For two random variables $\s{X}$ and $\s{Y}$, we write $\s{X}\indep\s{Y}$ if $\s{X}$ and $\s{Y}$ are statistically independent. We will frequently use standard big $O$ notations, and finally all logarithms are defined w.r.t. the natural basis.

The rest of the paper is organized as follows. In Section~\ref{sec:main}, we consider the problem of detecting and recovering the presence of a large general hidden subgraph under the subgraph ensemble. In particular, Subsection~\ref{subsec:stat} is devoted for the statistical limits; we derive present statistical lower and upper bound for these inference tasks, and discuss the obtained results. In Subsection~\ref{subsec:poly} we propose several polynomial-time algorithms and analyze their performance. Then, in Subsection~\ref{subsec:gaps} we provide a rigorous evidence for the statistical-computational gaps using the low-degree polynomial method. The proofs of our main results appear in Section~\ref{sec:proofs}, and finally, we discuss our main conclusions and outlook in Section~\ref{ref:conc_out}.

\section{Main Results}\label{sec:main}

\subsection{Statistical Limits}\label{subsec:stat}

In this subsection, we present the statistical limits of detecting and recovering the planted structure $\Gamma_k$ under the subgraph ensemble. We start with the detection boundary, giving sufficient conditions for the problem to be too hard for any test. To that end, we define:
\begin{align}
\s{D}_{\s{H}}&\triangleq\binom{n}{v(\s{H})}\frac{v(\s{H})!}{|\s{Aut}(\s{H})|}\p{\frac{q}{1-q}}^{e(\s{H})}(1-q)^{\binom{v(\s{H})}{2}},\label{cond:IT}
\end{align}
for any subgraph $\s{H}\subseteq\Gamma_k$ with $e(\s{H})>0$. As we mentioned in the Introduction, our focus in this subsection will be on the transition in the dense regime, when $q$ is near constant, namely, $n^{-o(1)}\leq q\leq1-n^{-o(1)}$. In Section~\ref{ref:conc_out}, we discuss other regimes of interest.
\begin{theorem}[Detection lower bound]\label{thm:IT_limit}
Consider the detection problem in \eqref{eqn:super_hypo}, for a fixed planted structure $\Gamma_k$ on $k$ vertices. Then, all tests are asymptotically powerless if $\min_{\s{H}\subseteq\Gamma_k:\;e(\s{H})>0}\s{D}_{\s{H}}\to\infty$.
\end{theorem}
Theorem~\ref{thm:IT_limit} holds for any structure $\Gamma_k$. For the following important family of subgraphs we can obtain a simpler representation of the statistical barrier. Specifically, we introduce the notion of graph maximum density. Denote by $d(\Gamma_k)\triangleq e(\Gamma_k)/v(\Gamma_k)$
the density of a graph $\Gamma_k$. Notice that $2\cdot d(\Gamma_k)$ is the average vertex degree in $\Gamma_k$. Define the \emph{maximum subgraph density} as \cite{bollobas_2001},
\begin{align}
m(\Gamma_k)\triangleq\max\ppp{d(\s{H}):\;\s{H}\subseteq\Gamma_k}.\label{eqn:maxDensity}
\end{align}
A graph $\Gamma_k$ is \emph{strictly balanced} if $d(\s{H})<d(\Gamma_k)$, for all proper subgraphs $\s{H}\subseteq\Gamma_k$. Note that trees, cycles, and cliques are strictly balanced graphs. 
\begin{corollary}[Strictly balanced subgraphs]\label{cor:1}
Consider the detection problem in Definition~\ref{def:detection}, and assume that $\Gamma_k$ is strictly balanced. Then, all tests are asymptotically powerless if $\s{D}_{\Gamma_k}\to\infty$.
\end{corollary}
The above results are, in fact, quite intuitive. Consider, for example, the case where $\Gamma_k$ is strictly balanced; in this case, $\s{D}_{\Gamma_k}$ characterizes the statistical limit. A little bit of thought reveals that $\s{D}_{\Gamma_k}$ is \emph{precisely} the expected number of copies of $\Gamma_k$ under the null hypothesis. Accordingly, in this case, $\s{D}_{\Gamma_k}\to\infty$ implies that, with high probability under the null, the $\Gamma_k$-number, i.e., the size of largest $\Gamma_k$-structure in $\s{G}$, is at least $k$, which is the size of the $\Gamma_k$-structure planted under the alternative and, therefore, detection is impossible. This is not enough to prove the result, however, as the $\Gamma_k$-number could still be even larger under the alternative; and even if this is not the case, it would only imply that the $\Gamma_k$-number test is powerless, but would not say anything about other tests. To prove Theorem~\ref{thm:IT_limit}, we adopt the standard approach based on studying the likelihood ratio test; see, for example, \cite[Chapter 8]{lehmann2005testing}. In this specific setting, the second moment method, which consists of showing that the variance of the likelihood ratio tends to zero suffices. For general planted structures, it turns out that the above intuition is not true anymore, and the optimal threshold depends on the density of the densest subgraph in $\Gamma_k$ in a non-trivial manner; specifically, note that $\min_{\s{H}\subseteq\Gamma_k:\;e(\s{H})>0}\s{D}_{\s{H}}$, characterizing the detection lower bound in Theorem~\ref{thm:IT_limit}, does \emph{not} necessarily equal to the expected number of copies $\s{D}_{\Gamma_k}$. Accordingly, the above (almost) folklore intuition, which holds for strictly balanced graphs (e.g., cliques), is imprecise for general structures. It should be emphasized that a similar phenomenon holds true for the number of copies of a given subgraph in $\calG(n,q)$, as mentioned in Subsection~\ref{subsec:litreature}. 

Computational considerations aside, the most natural test for detecting the presence of a $\Gamma_k$ structure is the $\Gamma_k$-number test given in Algorithm~\ref{algo:optAlg}. In the second step of this algorithm we search for the largest densest subgraph $\s{H}\subseteq\Gamma_k$ (in the sense of \eqref{cond:IT}. Note that it is also optimal to search for the largest structure $\Gamma_k$ in $\s{G}$ instead, however, in some cases, the computational complexity of the former can be smaller.
\begin{theorem}[Detection upper bound]\label{thm:IT_limit_alog}
Consider the detection problem in \eqref{eqn:super_hypo}, for a fixed planted structure $\Gamma_k$ on $k$ vertices. Algorithm~\ref{algo:optAlg} is asymptotically powerful if $\min_{\s{H}\subseteq\Gamma_k:\;e(\s{H})>0}\s{D}_{\s{H}}\to0$. Also, if $\Gamma_k$ is strictly balanced, then Algorithm~\ref{algo:optAlg} is asymptotically powerful if $\s{D}_{\Gamma_k}\to0$.
\end{theorem}
Establishing the statistical limits for the detection task, we now consider the recovery problem in Definition~\ref{def:recovery}. The following theorem states that over the subgraph ensemble, recovery is not more difficult than detection.
\begin{theorem}[Recovery upper \& lower bounds]\label{thm:IT_limit_alog_recovery}
Consider the detection problem in \eqref{eqn:super_hypo}, for a fixed planted structure $\Gamma_k$ on $k$ vertices. There is an algorithm that achieves asymptotic recovery if $\min_{\s{H}\subseteq\Gamma_k:\;e(\s{H})>0}\s{D}_{\s{H}}\to0$, while exact recovery is impossible if $\min_{\s{H}\subseteq\Gamma_k:\;e(\s{H})>0}\s{D}_{\s{H}}\to\infty$.
\end{theorem}
The recovery algorithm achieving the statistical bound in Theorem~\ref{thm:IT_limit_alog_recovery} outputs any subgraph $\Gamma_k$ of size $k$ in $\s{G}$ if such a subgraph exists, and the empty set otherwise. To conclude this subsection, we provide examples for the statistical barriers of a few simple structures, assuming that $q$ is fixed:
\begin{itemize}[leftmargin=*]
\itemsep-0.3em
\item \emph{Clique:} for a planted clique we obtain the folklore statistical barrier at $2\log_{\frac{1}{q}}n$.
\item \emph{Independent set:} for a planted independent set the statistical barrier is at $2\log_{\frac{1}{1-q}}n$.
\item \emph{Line graph:} for a planted line graph the statistical barrier is at $2\log_{\frac{1}{1-q}}\pp{n\frac{q}{1-q}}$.
\end{itemize}

\begin{algorithm}[t]
\caption{\texttt{Optimal Detection Algorithm}\label{algo:optAlg}}
\footnotesize
\begin{algorithmic}[1]
\REQUIRE Structure $\Gamma_k$, and a graph $\s{G}$.
\STATE Find $\hat{\s{H}} = \arg\min_{\s{H}\subseteq\Gamma_k:\;e(\s{H})>0}\s{D}_{\s{H}}$.
\STATE Find the largest $\hat{\s{H}}$ structure in $\s{G}$, and denote its size by $\s{M}$. 
\STATE If $\s{M}\geq v(\hat{\s{H}})$ decide $\calH_1$; otherwise, decide $\calH_0$. 
\end{algorithmic}
\end{algorithm}

\subsection{Computationally Efficient Detection Algorithms}\label{subsec:poly}

The statistical optimal algorithms developed in the previous subsection involve computing a sum of $\binom{n}{k}$ terms, which is clearly not computationally efficient. Below, we investigate what can be done in polynomial-time. For the rest of this paper we focus on the detection problem.

\vspace{0.2cm}
\noindent\textbf{Total degree test.} The simplest reasonable test one can imagine is the \emph{total degree test}, which rejects when the total number of edges in the graph is unusually large/small. Specifically, let $\s{W}(\s{G})$ be the total number of edges in the observed graph $\s{G}$. Below we assume that $q<1/2$, and then discuss the complementary case. Under the null hypothesis, it is clear that $\s{W}(\s{G})\sim\s{Binomial}\p{\binom{n}{2},q}$, while under the alternative hypothesis $\s{W}(\s{G})\sim e(\Gamma_k)+\s{Binomial}\p{\binom{n}{2}-\binom{k}{2},q}$. For the later, note that we constraint both the edges of $\Gamma_k$ and the edges of $\Gamma_k^c$ (w.r.t. the complete graph $\s{K}_k$ on $k$ vertices) to exist and not exist, respectively, and therefore the total number of possible random edges are $\binom{n}{2}-e(\Gamma_k)-e(\Gamma_k^c) = \binom{n}{2}-\binom{k}{2}$. Accordingly, simple statistical inference considerations suggest that, it is natural to define the test: decide $\phi_{\s{Tot}}(\s{G})=1$ iff
\begin{align}
\s{W}(\s{G})&\geq\frac{\bE_{\calH_0}[\s{W}(\s{G})]+\bE_{\calH_1}[\s{W}(\s{G})]}{2} \nonumber\\
&= q\cdot\binom{n}{2}+\frac{e(\Gamma_k)-q\cdot\binom{k}{2}}{2}\triangleq\s{W}^\star.\label{eqn:OnesidedTest}
\end{align}
However, one quickly realizes that the above test is problematic when $e(\Gamma_k)-q\cdot\binom{k}{2}$ is negative. The remedy is simple: when $e(\Gamma_k)-q\cdot\binom{k}{2}\geq0$ we apply the test in \eqref{eqn:OnesidedTest}, otherwise, we flip the decision, i.e., we decide $\phi_{\s{Tot}}(\s{G})=1$ iff $\s{W}(\s{G})<\s{W}^\star$. The intuition for this is that when $e(\Gamma_k)-q\cdot\binom{k}{2}\geq0$, the total number of edges under $\calH_1$ is unusually large, while when $e(\Gamma_k)-q\cdot\binom{k}{2}\leq0$, the total number of edges under $\calH_0$ is unusually large. Finally, if $q>1/2$, we follow the above procedure but replace $\s{G}$ with its graph-complementary $\s{G}^c$. To wit, for $q>1/2$ and $e(\Gamma_k^c)-(1-q)\cdot\binom{k}{2}\geq0$, decide $\phi_{\s{Tot}}(\s{G}^c)=1$ iff
\begin{align}
\s{W}(\s{G}^c)\geq (1-q)\cdot\binom{n}{2}+\frac{e(\Gamma_k^c)-(1-q)\cdot\binom{k}{2}}{2}
,\label{eqn:OnesidedTest2}
\end{align}
while for $q>1/2$ and $e(\Gamma_k^c)-(1-q)\cdot\binom{k}{2}\leq0$, we flip the decision. We refer to the combination of all these cases as the total degree test.
\begin{theorem}[Total degree test]\label{thm:4}
Let $\delta\in(0,1)$, and consider the total degree test $\phi_{\s{Tot}}$ defined above. Then, the average $\mathsf{Type}$ $\mathsf{I}$+$\mathsf{II}$ risk of $\phi_{\s{Tot}}$ is $\gamma_n (\phi_{\s{Tot}})\leq\delta$, if
\begin{align}
\frac{\abs{e(\Gamma_k)-q\cdot\binom{k}{2}}^2\pp{2\log\frac{2}{\delta}}^{-1/2}}{\min(q,1-q)\cdot\binom{n}{2}+\abs{e(\Gamma_k)-q\cdot\binom{k}{2}}}\geq 1.
\label{eqn:totalfinite}
\end{align}
\end{theorem}
\vspace{-0.2cm}
Not surprisingly, this test is considerably weaker than the essentially optimal test studied in the previous section, since the total degree test ignores any structure of the graph. Indeed, the gap between the statistical and computational limits we derived so far appear abysmal. Roughly speaking, for $q$ near constant, i.e., $n^{-o(1)}\leq q\leq1-n^{-o(1)}$, the total degree test succeeds w.h.p. if $k\geq\Omega(\sqrt{n\log n})$, \emph{for any structure $\Gamma_k$}. Note that Theorem~\ref{thm:4} holds for any choice of $q$; further implications are discussed in Subsection~\ref{subsec:further}.

\vspace{0.2cm}
\noindent\textbf{Spectral algorithm.} In what follows, we describe a spectral method that is able to shave off the logarithmic factor from the simplistic bound described above. Specifically, the \emph{spectral test} we study here is based on the spectral norm of the adjacency matrices of the $\s{G}$ and its \emph{complement} $\s{G}^c$. To present the main idea, it is more convenient to work with the adjacency matrix $\mathbf{A}$, given by:
\begin{align}
    \mathbf{A}_{ij} \triangleq \begin{cases}
    +1\ &\mathsf{if}\;i\neq j\;\mathsf{and}\;i\sim j\\
		0\ &\mathsf{otherwise,}
    \end{cases}
\end{align}
for any $1\leq i<j\leq n$, where ``$i\sim j$" means that vertices $i$ and $j$ are adjacent in $\s{G}$. Under the null hypothesis, it is clear that the entries of $\mathbf{A}$ are statistically independent and symmetric. Recall that the spectral norm of a real-valued symmetric matrix $\mathbf{A}$ is
\begin{align}
\norm{\mathbf{A}}_{\s{op}}\triangleq\sup_{\mathbf{x}\in\mathbb{S}^{n-1}}|\mathbf{x}^T\mathbf{A}\mathbf{x}|,
\end{align}
where the supremum is taken over the Euclidean unit sphere $\mathbb{S}^{n-1}=\{\mathbf{x}\in\mathbb{R}^n:\norm{\mathbf{x}}_2=1\}$. Typically, the spectral test amounts for comparing the spectral norm of the adjacency matrix to some threshold. The intuition behind this procedure is that in the presence of a sufficiently large hidden structure, $\norm{\mathbf{A}}_{\s{op}}$ is larger, with high probability, than that under the null hypothesis. However, in our case, this is not necessarily true. Indeed, under $\calH_1$ consider the unit vector $\mathbf{x}_{\Gamma_k}$ with entries $(x_1,\ldots,x_n)$ such that $x_i=1/\sqrt{k}$ if $i\in v(\Gamma_k)$, and $x_i=0$, otherwise. Then,
\begin{align}
\norm{\mathbf{A}}_{\s{op}}\geq |\mathbf{x}_{\Gamma_k}^T\mathbf{A}\mathbf{x}_{\Gamma_k}| = 2\cdot\frac{e(\Gamma_k)}{k}.\label{eqn:spectlower}
\end{align}
For cliques the r.h.s. of \eqref{eqn:spectlower} is proportional to $k$, and thus large, and can be compared to some threshold. However, if $\Gamma_k$ represents a line graph then $\frac{e(\Gamma_k)}{k}=\frac{k-1}{k}$, making the r.h.s. of \eqref{eqn:spectlower} almost independent of $k$. To overcome this issue, we look at $\norm{\mathbf{A}}_{\s{op}}+\norm{\mathbf{A}^c}_{\s{op}}$, where $\mathbf{A}^c$ corresponds to the adjacency matrix of $\s{G}^c$. Then, we have,
\begin{align}
&\norm{\mathbf{A}}_{\s{op}}+\norm{\mathbf{A}^c}_{\s{op}}\geq |\mathbf{x}_{\Gamma_k}^T\mathbf{A}\mathbf{x}_{\Gamma_k}|+|\mathbf{x}_{\Gamma_k}^T\mathbf{A}^c\mathbf{x}_{\Gamma_k}| \nonumber\\
&\quad\quad\quad\quad\quad\quad\ \ = 2\cdot\frac{e(\Gamma_k)}{k}+2\cdot\frac{e(\Gamma_k^c)}{k} = k-1.\label{eqn:spectlower2}
\end{align}
To present our main result, we define for any $\delta>0$,
\begin{align}
\varphi(n,q,\delta)\triangleq 4\sqrt{q(1-q)n}+\frac{2\sqrt{q(1-q)n}\log\frac{4n}{\delta^2}}{[q(1-q)n]^{1/6}-\frac{1}{2}\log\frac{4n}{\delta^2}},
\end{align}
and $\s{S}(\mathbf{A})\triangleq\norm{\mathbf{A}-\bE_{\calH_0}\mathbf{A}}_{\s{op}}+\norm{\mathbf{A}^c-\bE_{\calH_0}\mathbf{A}^c}_{\s{op}}$. Our spectral test $\phi_{\s{spec}}$ accepts the null hypothesis iff $\s{S}(\mathbf{A})\leq\varphi(n,q,\delta)$. We have the following result.
\begin{theorem}[Spectral test]\label{thm:5}
Let $\delta\in(0,1)$, and consider the spectral test $\phi_{\s{spec}}$. Then, the average $\mathsf{Type}$ $\mathsf{I}$+$\mathsf{II}$ risk is $\gamma_n (\phi_{\s{spec}})\leq\delta$, if $2|2e(\Gamma_k)/k - (k-1)q|\geq \varphi(n,q,\delta)$.
\end{theorem}
Note that the spectral norm of a matrix is computable in polynomial time and hence this test is computationally feasible though not as efficient as just counting edges. For $q$ near constant it can be seen that the spectral norm test succeeds w.h.p. if $k\geq\Omega(\sqrt{n})$, \emph{for any structure $\Gamma_k$}. 

\subsection{Statistical-Computational Gaps}\label{subsec:gaps}

\noindent\textbf{Basics of the low-degree method.} We start by giving a brief introduction to the low-degree polynomial method. The premise of this method is to take low-degree multivariate polynomials in the entries of the observations as a proxy for efficiently-computable functions. The ideas below were first developed in a sequence of works in the sum-of-squares optimization literature \cite{barak2016nearly,Hopkins18,hopkins2017bayesian,hopkins2017power}.

In the following, we follow the notations and definition in \cite{Hopkins18,Dmitriy19}. Any distribution $\pr_{\calH_0}$ on $\Omega_n=\{0,1\}^{\binom{n}{2}}$ induces an inner product of measurable functions $f,g:\Omega_n\to\mathbb{R}$ given by $\left\langle f,g \right\rangle_{\calH_0} = \bE_{\calH_0}[f(\s{G})g(\s{G})]$, and norm $\norm{f}_{\calH_0} = \left\langle f,f \right\rangle_{\calH_0}^{1/2}$. We Let $L^2(\pr_{\calH_0})$ denote the Hilbert space consisting of functions $f$ for which $\norm{f}_{\calH_0}<\infty$, endowed with the above inner product and norm. In the computationally-unbounded case, the Neyman-Pearson lemma shows that the likelihood ratio test achieves the optimal tradeoff between $\mathsf{Type}$-$\mathsf{I}$ and $\mathsf{Type}$-$\mathsf{II}$ error probabilities. Furthermore, it is well-known that the same test optimally distinguishes $\pr_{\calH_0}$ from $\pr_{\calH_1}$ in the $L^2$ sense. In fact, denoting by $\s{L}_n\triangleq\pr_{\calH_1}/\pr_{\calH_0}$ the likelihood ratio, then the second moment method for contiguity shows that if $\norm{\s{L}_n}_{\calH_0}^2$ remains bounded as $n\to\infty$, then $\pr_{\calH_1}$ is contiguous to $\pr_{\calH_0}$. This implies that $\pr_{\calH_1}$ and $\pr_{\calH_0}$ are statistically indistinguishable, i.e., no test can have both $\mathsf{Type}$-$\mathsf{I}$ and $\mathsf{Type}$-$\mathsf{II}$ error probabilities tending to zero. 

We now describe the low-degree method. The idea is to find the low-degree polynomial that best distinguishes $\pr_{\calH_0}$ from $\pr_{\calH_1}$ in the $L^2$ sense. To that end, we let $\calV_{n,\leq\s{D}}\subset L^2(\pr_{\calH_0})$ denote the linear subspace of polynomials $\Omega_n\to\mathbb{R}$ of degree at most $\s{D}\in\mathbb{N}$. We define further $\calP_{\leq \s{D}}: L^2(\pr_{\calH_0})\to\calV_{n,\leq\s{D}}$ the orthogonal projection operator. Then, the \emph{$\s{D}$-low-degree likelihood ratio} $\s{L}_{n,\leq \s{D}}$ is the projection of a function $\s{L}_{n}$ to the span of coordinate-degree-$\s{D}$ functions, where the projection is orthogonal with respect to the inner product $\left\langle \cdot,\cdot \right\rangle_{\calH_0}$. As discussed above, the likelihood ratio optimally distinguishes $\pr_{\calH_0}$ from $\pr_{\calH_1}$ in the $L^2$ sense. The next lemma shows that over the set of low-degree polynomials, the $\s{D}$-low-degree likelihood ratio have the exhibit the same property.
\begin{lemma}[Optimally of $\s{L}_{n,\leq \s{D}}$ {\cite{hopkins2017bayesian,hopkins2017power,Dmitriy19}}]\label{lem:Dmitriy}
Consider the following optimization problem:
\begin{equation}
\begin{aligned}
\mathrm{max}
\;\bE_{\calH_1}f(\s{G})
\quad\mathrm{s.t.}
\quad\bE_{\calH_0}f^2(\s{G}) = 1,\; f\in\calV_{n,\leq\s{D}},
\end{aligned}\label{eqn:optimizationProblem}
\end{equation}
Then, the unique solution $f^\star$ for \eqref{eqn:optimizationProblem} is the $\s{D}$-low degree likelihood ratio $f^\star = \s{L}_{n,\leq \s{D}}/\norm{\s{L}_{n,\leq \s{D}}}_{\calH_0}$, and the value of the optimization problem is $\norm{\s{L}_{n,\leq \s{D}}}_{\calH_0}$. 
\end{lemma}
\vspace{-0.2cm}
As was mentioned above, in the computationally-unbounded regime, an important property of the likelihood ratio is that if $\norm{\s{L}_n}_{\calH_0}$ is bounded then $\pr_{\calH_0}$ and $\pr_{\calH_1}$ are statistically indistinguishable. The following conjecture states that a computational analogue of this property holds, with $\s{L}_{n,\leq \s{D}}$ playing the role of the likelihood ratio. In fact it also postulates that polynomials of degree $\approx\log n$ are a proxy for polynomial-time algorithms. The conjecture below is based on \cite{Hopkins18,hopkins2017bayesian,hopkins2017power}, and \cite[Conj. 2.2.4]{Hopkins18}. We give an informal statement of this conjecture which appears in \cite[Conj. 1.16]{Dmitriy19}. For a precise statement, we refer the reader to, e.g., \cite[Conj. 2.2.4]{Hopkins18} and \cite[Sec. 4]{Dmitriy19}.
\begin{conjecture}[Low-degree conj., informal]\label{conj:1}
Given a sequence of probability measures $\pr_{\calH_0}$ and $\pr_{\calH_1}$, if there exists $\epsilon>0$ and $\s{D} = \s{D}(n)\geq (\log n)^{1+\epsilon}$, such that $\norm{\s{L}_{n,\leq \s{D}}}_{\calH_0}$ remains bounded as $n\to\infty$, then there is no polynomial-time algorithm that distinguishes $\pr_{\calH_0}$ and $\pr_{\calH_1}$.
\end{conjecture}
\vspace{-0.2cm}
In the sequel, we will rely on Conjecture~\ref{conj:1} to give an evidence for the statistical-computational gap observed in the previous section. At this point we would like to mention \cite[Hypothesis 2.1.5]{Hopkins18}, which states a more general form of Conjecture~\ref{conj:1} in the sense that it postulates that degree-$\s{D}$ polynomials are a proxy for $n^{O(D)}$-time algorithms. Note that if $\norm{\s{L}_{n,\leq \s{D}}}_{\calH_0} = O(1)$, then we expect strong detection in time $\s{T}(n) = e^{\s{D}(n)}$ to be impossible.

\vspace{0.2cm}
\noindent\textbf{Gaps in the subgraph ensemble.} We are now in a position to state our main result of this subsection. 
\begin{theorem}[Statistical-computational gap]\label{thm:6}
Consider the detection problem in \eqref{eqn:super_hypo} and assume that $n^{-o(1)}\leq q\leq 1-n^{-o(1)}$. Then, for any planted structure $\Gamma_k$, and every $\epsilon>0$, if $k=n^{1/2-\epsilon}$, then $\norm{\s{L}_{n,\leq \s{D}}}_{\calH_0}\leq O(1)$, for any $\s{D} = \Omega(\log n)$.
\end{theorem}
Theorem~\ref{thm:6} implies that if we take degree-$\log n$ polynomials as a proxy for all efficient algorithms, our calculations predict that an $n^{O(\log n)}$ algorithm does not exist when $k\ll\sqrt{n}$. These predictions agree precisely with the previously established statistical-computational tradeoffs in the previous subsections. A more explicit formula for the computational barrier which exhibits dependency on $\s{D}$ and $q$ can be deduced from the proof of Theorem~\ref{thm:6}; to keep the exposition simple we opted to present the refined result above. 

\subsection{Further Discussion}\label{subsec:further}

The previous subsections characterizes the statistical and computational barriers tightly, when $q$ is near constant. However, it is important to understand the dependency of these barriers for other scalings of $q$; in particular, when $q$ is polynomially small $q = \Theta(n^{-\alpha})$, and when $q$ is very close to unity $q = 1-\Theta(n^{-\alpha})$, for $\alpha>0$. For example, in case of cliques, it is well-known that the statistical barrier is $\frac{1-q}{q}\ll\min(k^{-1},n^{2}k^{-4})$, while for independent sets the barrier is $\frac{q}{1-q}\ll\min(k^{-1},n^{2}k^{-4})$. Our lower bound in Theorem~\ref{thm:IT_limit} captures the first term in the minimum. The statistical optimal algorithm in this case is a combination of the total degree test and a scan test which finds the densest $k\times k$ subgraph in $\s{G}$. As mentioned above, Theorems~\ref{thm:4} and \ref{thm:5} hold true for any choice of $q$. Accordingly, for cliques, if $q = 1-\Theta(n^{-\alpha})$, then Theorem~\ref{thm:4} implies that the total degree test is powerful at the shifted threshold $k^2 = \Theta(n^{1+\alpha/2})$, which agree with the above statistical barrier. For independent set the situation is, of course, flipped. For line graphs, it can be shown that the total degree test barrier is, up to polylog-factors, consistent with that of independent set. On the other hand, for cliques, if $q = \Theta(n^{-\alpha})$ then the optimal detection test in Subsection~\ref{subsec:stat} is in fact efficient, and the problem begins to be easy when $k = \Theta(1)$. This barrier is captured by Theorem~\ref{thm:IT_limit}. Finally, note that for these scalings of $q$, the spectral test is inferior compared to the total degree test. In terms of the statistical-computational gaps, we would like to mention that while the statement of Theorem~\ref{thm:6} assumes that $q$ is near constant, our proof sheds light on other scaling of $q$. In particular, the proof of Theorem~\ref{thm:6} gives further evidence for the computational barrier of cliques at $k^2=\Theta(n^{1+\alpha/2})$ for $q = 1-\Theta(n^{-\alpha})$, proved in \cite{brennan18a} using the technique of average-case reductions from the planted clique conjecture. The same holds for independent sets in the complementary regime. For other structures the exact dependency of the computational barrier on $q$ and $\Gamma_k$ is currently unknown (see, Section~\ref{ref:conc_out}). 

\section{Proofs}\label{sec:proofs}

\subsection{Proof of Theorem~\ref{thm:IT_limit}}

For a test $\varphi:\ppp{0,1}^{n\times n}\to\{0,1\}$ the average-case Type I+II error probability is given by
\begin{align}
    \gamma_n(\varphi)\triangleq\pr_{\calH_0}(\varphi(\s{G})=1)+\pr_{\calH_1}(\varphi(\s{G})=0).
\end{align}
To lower bound the above quantity for any test, we use the fact that the likelihood ratio test minimizes the average risk. In our case, the likelihood ratio is given as follows,
\begin{align}
\s{L}(\s{G}) = \frac{\pr_{\calH_1}(\s{G})}{\pr_{\calH_0}(\s{G})}.
\end{align}
Then, it is well-known fact that the test $\varphi(\s{G}) = \Ind\pp{\s{L}(\s{G})>1}$ minimizes the average risk, with risk given by
\begin{align}
\gamma_{\s{opt}} = \pr_{\calH_0}(\s{L}(\s{G})>1)+\pr_{\calH_0}\pp{\s{L}(\s{G})\cdot\Ind\pp{\s{L}(\s{G})\leq1}}.
\end{align}
Therefore, it suffices to show that $\gamma_{\s{opt}}\to1$. To prove Theorem~\ref{thm:IT_limit} we use the second moment argument. Specifically, using Cauchy-Schwarz inequality, we have
\begin{align}
\gamma_{\s{opt}} = 1-\frac{1}{2}\bE_{\calH_0}|\s{L}(\s{G})-1|\geq 1-\frac{1}{2}\sqrt{\s{var}_0(\s{L}(\s{G}))}.
\end{align}
Hence, it suffices to prove that $\s{var}_0(\s{L}(\s{G}))\to0$ under the theorem conditions. To that end, we show that $\bE_{\calH_0}[\s{L}^2(\s{G})]\leq1+o(1)$. Below, without loss of generality we assume that $q<1/2$; the complementary case follows by analyzing the graph complement $\s{G}^c$.

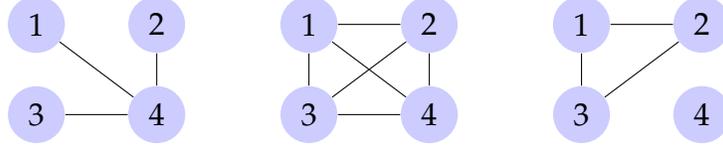
\begin{figure}[t!]
\centering
\begin{tikzpicture}
  [scale=.4,auto=left,every node/.style={circle,fill=blue!20}]
  \node (n6) at (13,9) {2};
  \node (n5) at (9,9)  {1};
  \node (n1) at (13,6) {4};
  \node (n2) at (9,6)  {3};

  \foreach \from/\to in {n5/n1,n1/n2,n1/n6}
    \draw (\from) -- (\to);

\end{tikzpicture}~~~~~~~~~~~
\begin{tikzpicture}
  [scale=.4,auto=left,every node/.style={circle,fill=blue!20}]
  \node (n6) at (13,9) {2};
  \node (n5) at (9,9)  {1};
  \node (n1) at (13,6) {4};
  \node (n2) at (9,6)  {3}; 

  \foreach \from/\to in {n5/n1,n5/n2,n5/n6,n1/n2,n1/n6,n2/n6}
    \draw (\from) -- (\to);
\end{tikzpicture}~~~~~~~~~~~
\begin{tikzpicture}
  [scale=.4,auto=left,every node/.style={circle,fill=blue!20}]
  \node (n6) at (13,9) {2};
  \node (n5) at (9,9)  {1};
  \node (n1) at (13,6) {4};
  \node (n2) at (9,6)  {3};

  \foreach \from/\to in {n5/n2,n5/n6,n2/n6}
    \draw (\from) -- (\to);

\end{tikzpicture}
\caption{Structure $\bar{\Gamma}_\ell$ (left), Complete graph $\calK_{\ell,4}$ (middle), Graph complement $\bar{\Gamma}_\ell^c$ (right).}
\label{fig:3}
\end{figure}

Next, let us compute the likelihood. To that end, we let $\calG_{n,k}$ be the set of all possible $\Gamma_k$--structures in the complete graph $\s{K}_n$ on $n$ nodes. Specifically, $\calG_{n,k}$ is the set of all possible subgraph copies $\bar{\Gamma}_1,\bar{\Gamma}_2,\ldots,\bar{\Gamma}_{|\calG_{n,k}|}$ of $\Gamma_k$ in $\s{K}_n$. A simple counting argument shows that $|\calG_{n,k}| = \binom{n}{k}\cdot\frac{k!}{|\s{Aut}(\Gamma_k)|}$, where $\s{Aut}(\Gamma_k)$ is the automorphism group of $\Gamma_k$. Let $e(\s{G})$ denote the number of edges in $\s{G}$. Note that $e(\bar{\Gamma}_\ell)=e(\Gamma_k)$, for any $\ell\in[|\calG_{n,k}|]$. Also, in the sequel, for any $\ell\in[|\calG_{n,k}|]$, we let $\bar{\Gamma}_\ell^c$ be the graph-complement of $\bar{\Gamma}_{\ell}$, where the complement is with respect to the $\bar{\Gamma}_{\ell}$ itself. Namely, if we let $\calK_{\ell,k}$ be the complete graph on the same $k$ labeled vertices of $\bar{\Gamma}_{\ell}$, then $\bar{\Gamma}_{\ell}^c = \calK_{\ell,k}\setminus\bar{\Gamma}_{\ell}$. Fig.~\ref{fig:3} gives an example. Then, for any graph $\s{g}\in\ppp{0,1}^{n\times n}$, we get:
\begin{align}
\pr_{\calH_1}(\s{G}=\s{g}) &= \frac{1}{|\calG_{n,k}|}\sum_{\ell=1}^{|\calG_{n,k}|}\pr_{\calH_0}(\s{G}=\s{g}\vert \bar{\Gamma}_\ell\in\s{G})\\
& = \frac{1}{|\calG_{n,k}|}\sum_{\ell=1}^{|\calG_{n,k}|}q^{e(\s{g})-e(\bar{\Gamma}_\ell)}(1-q)^{\binom{n}{2}-e(\s{g})-e(\bar{\Gamma}_\ell^c)}\Ind\pp{\bar{\Gamma}_\ell\in\s{g}}\\
& = \frac{q^{-e(\Gamma_k)}(1-q)^{-e(\Gamma_k^c)}}{|\calG_{n,k}|}q^{e(\s{g})}(1-q)^{\binom{n}{2}-e(\s{g})}\sum_{\ell=1}^{|\calG_{n,k}|}\Ind\pp{\bar{\Gamma}_\ell\in\s{g}}\\
& = \frac{q^{-e(\Gamma_k)}(1-q)^{-e(\Gamma_k^c)}}{|\calG_{n,k}|}\pr_{\calH_0}(\s{G}=\s{g})\sum_{\ell=1}^{|\calG_{n,k}|}\Ind\pp{\bar{\Gamma}_\ell\in\s{g}}\\
& = \frac{\calN_{\Gamma_k}}{\bE_{\calH_0}(\calN_{\Gamma_k})}\cdot\pr_{\calH_0}(\s{G}=\s{g}),
\end{align}
where $\calN_{\Gamma_k}\triangleq \sum_{\ell=1}^{|\calG_{n,k}|}\Ind\pp{\bar{\Gamma}_\ell\in\s{g}}$, and we have used the fact that
\begin{align}
\bE_{\calH_0}(\calN_{\Gamma_k}) &= |\calG_{n,k}|\cdot q^{e(\Gamma_k)}(1-q)^{e(\Gamma_k^c)} \\
&= |\calG_{n,k}|\cdot \p{\frac{q}{1-q}}^{e(\Gamma_k)}(1-q)^{\binom{k}{2}}.
\end{align}
Therefore, we obtain that $\s{L}(\s{G}) = \calN_{\Gamma_k}/\bE_{\calH_0}(\calN_{\Gamma_k})$, which is the observed number of $\Gamma_k$--structures of size $k$ divided by their expected number under the null hypothesis. 

Let us analyze the second moment of the likelihood function. We start with the case where $\Gamma_k$ is strictly balanced, and then move forward to the general case. We will start by upper bounding $\bE_{\calH_0}[\calN_{\Gamma_k}^2]$. First, note that
\begin{align}
\bE_{\calH_0}[\calN_{\Gamma_k}^2] &= \sum_{\ell_1=1}^{|\calG_{n,k}|}\sum_{\ell_2=1}^{|\calG_{n,k}|}\pr_{\calH_0}\pp{\bar{\Gamma}_{\ell_1},\bar{\Gamma}_{\ell_2}\in\s{G}}.
\end{align}
We need to evaluate the probability both  $\bar{\Gamma}_{\ell_1}$ and $\bar{\Gamma}_{\ell_2}$ appear as induced subgraphs in $\s{G}$. However, some pairs cannot coexist together: if $\bar{\Gamma}_{\ell_1}^c\cap\bar{\Gamma}_{\ell_2}\neq\emptyset$ or $\bar{\Gamma}_{\ell_2}^c\cap\bar{\Gamma}_{\ell_1}\neq\emptyset$, then both structures cannot appear in $\s{G}$ simultaneously. Indeed, in those cases there are edges that are not allowed to appear in one graph (so that it will appear as an induced subgraph of $\s{G}$), but must appear in the other graph. Accordingly, let $\calL$ denote the set of possible pairs $(\bar{\Gamma}_{\ell_1},\bar{\Gamma}_{\ell_2})$. Then, for any $(\bar{\Gamma}_{\ell_1},\bar{\Gamma}_{\ell_2})\in\calL$, it is clear that \begin{align}
\pr_{\calH_0}\pp{\bar{\Gamma}_{\ell_1},\bar{\Gamma}_{\ell_2}\in\s{G}} = q^{e(\bar{\Gamma}_{\ell_1}\cup\bar{\Gamma}_{\ell_2})}(1-q)^{e(\bar{\Gamma}_{\ell_1}^c\cup\bar{\Gamma}_{\ell_2}^c)}.
\end{align} 
Now, note that 
\begin{align}
\bar{\Gamma}_{\ell_1}^c\cap\bar{\Gamma}_{\ell_2}^c &= (\calK_{\ell_1,k}\setminus\bar{\Gamma}_{\ell_1})\cap(\calK_{\ell_2,k}\setminus\bar{\Gamma}_{\ell_2})\\
& = [\calK_{\ell_1,k}\cap(\calK_{\ell_2,k}\setminus\bar{\Gamma}_{\ell_2})]\setminus[(\bar{\Gamma}_{\ell_1}\cap\calK_{\ell_2,k})\setminus\bar{\Gamma}_{\ell_1}\cap\bar{\Gamma}_{\ell_2}]\\
& = [\calK_{\ell_1,k}\cap\calK_{\ell_2,k}\setminus\calK_{\ell_1,k}\cap\bar{\Gamma}_{\ell_2})]\setminus[(\bar{\Gamma}_{\ell_1}\cap\calK_{\ell_2,k})\setminus\bar{\Gamma}_{\ell_1}\cap\bar{\Gamma}_{\ell_2}],
\end{align}
and therefore,
\begin{align}
e(\bar{\Gamma}_{\ell_1}^c\cap\bar{\Gamma}_{\ell_2}^c) &= e(\calK_{\ell_1,k}\cap\calK_{\ell_2,k})-e(\calK_{\ell_1,k}\cap\bar{\Gamma}_{\ell_2}))-e(\calK_{\ell_2,k}\cap\bar{\Gamma}_{\ell_1})+e(\bar{\Gamma}_{\ell_1}\cap\bar{\Gamma}_{\ell_2})\\
& = e(\calK_{\ell_1,k}\cap\calK_{\ell_2,k})-e(\bar{\Gamma}_{\ell_1}\cap\bar{\Gamma}_{\ell_2}),
\end{align}
where in the last equality we have used the fact that $e(\calK_{\ell_1,k}\cap\bar{\Gamma}_{\ell_2})=e(\calK_{\ell_2,k}\cap\bar{\Gamma}_{\ell_1})=e(\bar{\Gamma}_{\ell_1}\cap\bar{\Gamma}_{\ell_2})$. Let $v(\s{G})$ denote the number of vertices in $\s{G}$. Then, since $\calK_{\ell_1,k}\cap\calK_{\ell_2,k}$ is a complete graph on $v(\bar{\Gamma}_{\ell_1}\cap\bar{\Gamma}_{\ell_2})$ vertices, we have,
\begin{align}
e(\bar{\Gamma}_{\ell_1}^c\cap\bar{\Gamma}_{\ell_2}^c) &=\binom{v(\bar{\Gamma}_{\ell_1}\cap\bar{\Gamma}_{\ell_2})}{2}-e(\bar{\Gamma}_{\ell_1}\cap\bar{\Gamma}_{\ell_2}).
\end{align}
The above coupled with the inclusion-exclusion principle imply that, 
\begin{align}
e(\bar{\Gamma}_{\ell_1}^c\cup\bar{\Gamma}_{\ell_2}^c) &= e(\bar{\Gamma}_{\ell_1}^c)+e(\bar{\Gamma}_{\ell_2}^c)-\binom{v(\bar{\Gamma}_{\ell_1}\cap\bar{\Gamma}_{\ell_2})}{2}+e(\bar{\Gamma}_{\ell_1}\cap\bar{\Gamma}_{\ell_2})\\
& = 2\cdot\binom{k}{2}-e(\bar{\Gamma}_{\ell_1}\cup\bar{\Gamma}_{\ell_2})-\binom{v(\bar{\Gamma}_{\ell_1}\cap\bar{\Gamma}_{\ell_2})}{2}.
\end{align}
Therefore,
\begin{align}
\bE_{\calH_0}[\calN_{\Gamma_k}^2]&= \sum_{(\bar{\Gamma}_{\ell_1},\bar{\Gamma}_{\ell_2})\in\calL}\p{\frac{q}{1-q}}^{e(\bar{\Gamma}_{\ell_1}\cup\bar{\Gamma}_{\ell_2})}(1-q)^{2\cdot\binom{k}{2}-\binom{v(\bar{\Gamma}_{\ell_1}\cap\bar{\Gamma}_{\ell_2})}{2}}\\
&\leq \sum_{\ell_1=1}^{|\calG_{n,k}|}\sum_{\ell_2=1}^{|\calG_{n,k}|}\p{\frac{q}{1-q}}^{e(\bar{\Gamma}_{\ell_1}\cup\bar{\Gamma}_{\ell_2})}(1-q)^{2\cdot\binom{k}{2}-\binom{v(\bar{\Gamma}_{\ell_1}\cap\bar{\Gamma}_{\ell_2})}{2}},\label{eqn:second_moment_strict}
\end{align}
and accordingly,
\begin{align}
\bE_{\calH_0}[\s{L}^2(\s{G})]&\leq \frac{1}{|\calG_{n,k}|^2}\sum_{\ell_1=1}^{|\calG_{n,k}|}\sum_{\ell_2=1}^{|\calG_{n,k}|}\frac{\p{\frac{q}{1-q}}^{e(\bar{\Gamma}_{\ell_1}\cup\bar{\Gamma}_{\ell_2})}(1-q)^{2\cdot\binom{k}{2}-\binom{v(\bar{\Gamma}_{\ell_1}\cap\bar{\Gamma}_{\ell_2})}{2}}}{q^{2e(\Gamma_k)}(1-q)^{2e(\Gamma_k^c)}}\label{eqn:summand}\\
&\triangleq \frac{1}{|\calG_{n,k}|^2}\sum_{\ell_1=1}^{|\calG_{n,k}|}\sum_{\ell_2=1}^{|\calG_{n,k}|}f(\bar{\Gamma}_{\ell_1}\cap\bar{\Gamma}_{\ell_2}),
\end{align}
where in the last equality we have used the facts that $e(\Gamma_k) = e(\bar{\Gamma}_{\ell_1}) = e(\bar{\Gamma}_{\ell_2})$, and that the summand in \eqref{eqn:summand} depends on $\bar{\Gamma}_{\ell_1}$ and $\bar{\Gamma}_{\ell_2}$ through $\bar{\Gamma}_{\ell_1}\cap\bar{\Gamma}_{\ell_2}$ only. Let $\pi$ be the uniform probability measure over $\calG_{n,k}$. Then,
\begin{align}
\bE_{\calH_0}[\s{L}^2(\s{G})] &= \bE_{\s{L}_1\indep\s{L}_2\sim\pi}\pp{f(\bar{\Gamma}_{\s{L}_1}\cap\bar{\Gamma}_{\s{L}_2})}\\
& = \bE_{\s{L}_1\indep\s{L}_2\sim\pi}\pp{f(\bar{\Gamma}_{\s{L}_1}\cap\bar{\Gamma}_{\s{L}_2})\Ind\p{e(\bar{\Gamma}_{\s{L}_1}\cap\bar{\Gamma}_{\s{L}_2})=0}}\nonumber\\
&\quad+\bE_{\s{L}_1\indep\s{L}_2\sim\pi}\pp{f(\bar{\Gamma}_{\s{L}_1}\cap\bar{\Gamma}_{\s{L}_2})\Ind\p{e(\bar{\Gamma}_{\s{L}_1}\cap\bar{\Gamma}_{\s{L}_2})>0}}.
\end{align}
Over the event $\{e(\bar{\Gamma}_{\s{L}_1}\cap\bar{\Gamma}_{\s{L}_2})=0\}$ we have $f(\bar{\Gamma}_{\s{L}_1}\cap\bar{\Gamma}_{\s{L}_2})=1$. Thus,
\begin{align}
\bE_{\calH_0}[\s{L}^2(\s{G})] & = \pr_{\s{L}_1\indep\s{L}_2\sim\pi}\pp{e(\bar{\Gamma}_{\s{L}_1}\cap\bar{\Gamma}_{\s{L}_2})=0}\nonumber\\
&\quad+\bE_{\s{L}_1\indep\s{L}_2\sim\pi}\pp{f(\bar{\Gamma}_{\s{L}_1}\cap\bar{\Gamma}_{\s{L}_2})\Ind\p{e(\bar{\Gamma}_{\s{L}_1}\cap\bar{\Gamma}_{\s{L}_2})>0}}\\
&\leq 1+\bE_{\s{L}_1\indep\s{L}_2\sim\pi}\pp{f(\bar{\Gamma}_{\s{L}_1}\cap\bar{\Gamma}_{\s{L}_2})\Ind\p{e(\bar{\Gamma}_{\s{L}_1}\cap\bar{\Gamma}_{\s{L}_2})>0}}.
\end{align}
Accordingly, we have
\begin{align}
\bE_{\calH_0}[\s{L}^2(\s{G})] &\leq 1+ \frac{1}{|\calG_{n,k}|^2}\sum_{(\ell_1,\ell_2):\;e(\bar{\Gamma}_{\ell_1}\cap\bar{\Gamma}_{\ell_2})>0}f(\bar{\Gamma}_{\ell_1}\cap\bar{\Gamma}_{\ell_2})\\
& = 1+\frac{1}{\pp{\bE_{\calH_0}(\calN_{\Gamma_k})}^2}\sum_{(\ell_1,\ell_2):\;e(\bar{\Gamma}_{\ell_1}\cap\bar{\Gamma}_{\ell_2})>0}\p{\frac{q}{1-q}}^{e(\bar{\Gamma}_{\ell_1}\cup\bar{\Gamma}_{\ell_2})}(1-q)^{2\cdot\binom{k}{2}-\binom{v(\bar{\Gamma}_{\ell_1}\cap\bar{\Gamma}_{\ell_2})}{2}}.\label{eqn:termRHS}
\end{align}
Next, we analyze the term on the r.h.s. of \eqref{eqn:termRHS}. To that end, we use the fact that $\Gamma_k$ is a strictly balanced graph. Suppose that $\bar{\Gamma}_{\ell_1}$ and $\bar{\Gamma}_{\ell_2}$ are isomorphic $\Gamma_k$, namely, $\bar{\Gamma}_{\ell_1},\bar{\Gamma}_{\ell_2}\simeq\Gamma_k$, and exactly $t\geq1$ vertices of $\bar{\Gamma}_{\ell_2}$ lie outside $v(\bar{\Gamma}_{\ell_1})$. Then, for $0<t<k$, we have
\begin{align}
e(\bar{\Gamma}_{\ell_1}\cup\bar{\Gamma}_{\ell_2})\geq e(\bar{\Gamma}_{\ell_1})+\frac{te(\bar{\Gamma}_{\ell_1})}{k}+\frac{1}{k} = \frac{(k+t)e(\Gamma_k)}{k}+\frac{1}{k}.
\end{align}
Indeed, since $\Gamma_k$ is strictly balanced, fewer than $(k-t)e(\bar{\Gamma}_{\ell_1})/k$ edges of $\bar{\Gamma}_{\ell_2}$ join
its $k-t$ vertices in $v(\bar{\Gamma}_{\ell_1}\cap\bar{\Gamma}_{\ell_2})$. Note that under the parameterization $s=k+t$, we have $s = v(\bar{\Gamma}_{\ell_1}\cup\bar{\Gamma}_{\ell_2})$, and since in \eqref{eqn:termRHS} we consider all pairs with at least one common edge, we have $k\leq s\leq 2k-2$. Grouping the summation term at the r.h.s. of \eqref{eqn:termRHS} by $v(\bar{\Gamma}_{\ell_1}\cup\bar{\Gamma}_{\ell_2})$, we get
\begin{align}
&\sum_{(\ell_1,\ell_2):\;e(\bar{\Gamma}_{\ell_1}\cap\bar{\Gamma}_{\ell_2})>0}\p{\frac{q}{1-q}}^{e(\bar{\Gamma}_{\ell_1}\cup\bar{\Gamma}_{\ell_2})}(1-q)^{2\cdot\binom{k}{2}-\binom{v(\bar{\Gamma}_{\ell_1}\cap\bar{\Gamma}_{\ell_2})}{2}} \nonumber\\
&\quad\quad\quad\leq \sum_{s=k}^{2k-2}\p{\frac{q}{1-q}}^{[se(\Gamma_k)+1]/k}(1-q)^{2\cdot\binom{k}{2}-\binom{2k-s}{2}}\cdot\abs{(\ell_1,\ell_2)\in\calG_{n,k}^2:\;v(\bar{\Gamma}_{\ell_1}\cup\bar{\Gamma}_{\ell_2})=s}.
\end{align}
In the above, we count the number of pairs $(\bar{\Gamma}_{\ell_1},\bar{\Gamma}_{\ell_2})$ of copies of $\Gamma_k$ such that $v(\bar{\Gamma}_{\ell_1}\cup\bar{\Gamma}_{\ell_2})=s$, or, equivalently, $v(\bar{\Gamma}_{\ell_1}\cap\bar{\Gamma}_{\ell_2})=2k-s$. We can upper bound this quantity by
\begin{align}
\abs{(\ell_1,\ell_2)\in\calG_{n,k}^2:\;v(\bar{\Gamma}_{\ell_1}\cup\bar{\Gamma}_{\ell_2})=s}\leq |\calG_{n,k}|\binom{k}{2k-s}|\calG_{n-k,s-k}|,
\end{align}
where the equality follows from the following reasoning: first, there are $|\calG_{n,k}|$ ways of picking a graph $\bar{\Gamma}_{\ell_1}$ of $k$ vertices from a graph on $n$ vertices. Then, for each such graph $\bar{\Gamma}_{\ell_1}$, there are exactly $\binom{k}{2k-s}$ ways to pick $2k-s$ nodes from $\bar{\Gamma}_{\ell_1}$ that will also be part of another graph $\bar{\Gamma}_{\ell_2}$. Once $\bar{\Gamma}_{\ell_1}$ and the nodes of $\bar{\Gamma}_{\ell_2}$ that will be shared with $\bar{\Gamma}_{\ell_1}$ have been determined, it remains to pick from $\bar{\Gamma}_{\ell_1}^c$ the remaining $k-(2k-s) = s-k$ nodes of $\bar{\Gamma}_{\ell_2}$, and there are exactly $|\calG_{n-k,s-k}|$ ways of doing that. Therefore,
\begin{align}
&\sum_{s=k}^{2k-2}\p{\frac{q}{1-q}}^{[se(\Gamma_k)+1]/k}(1-q)^{2\cdot\binom{k}{2}-\binom{2k-s}{2}}\cdot\abs{(\ell_1,\ell_2)\in\calG_{n,k}^2:\;v(\bar{\Gamma}_{\ell_1}\cup\bar{\Gamma}_{\ell_2})=s}\nonumber\\
&\quad\quad \leq|\calG_{n,k}|(1-q)^{2\binom{k}{2}}\cdot\sum_{s=k}^{2k-2}\p{\frac{q}{1-q}}^{se(\Gamma_k)/k}(1-q)^{-\binom{2k-s}{2}}\binom{k}{2k-s}\cdot|\calG_{n-k,s-k}|\\
&\quad\quad\stackrel{u\leftarrow2k-s}{=}|\calG_{n,k}|\p{\frac{q}{1-q}}^{2e(\Gamma_k)}(1-q)^{2\binom{k}{2}}\nonumber\\
&\quad\quad\quad\quad\quad\quad\cdot\sum_{u=2}^k\binom{k}{u}\p{\frac{q}{1-q}}^{-ue(\Gamma_k)/k}(1-q)^{-\binom{u}{2}}\cdot|\calG_{n-k,k-u}|\\
&\quad\quad = |\calG_{n,k}|\p{\frac{q}{1-q}}^{2e(\Gamma_k)}(1-q)^{2\binom{k}{2}}\cdot\sum_{u=2}^k\binom{k}{u}\p{\frac{q}{1-q}}^{-ue(\Gamma_k)/k}(1-q)^{-\binom{u}{2}}\cdot|\calG_{n-k,k-u}|\\
&\quad\quad = \pp{\bE_{\calH_0}(\calN_{\Gamma_k})}^2\cdot\sum_{u=2}^k\binom{k}{u}\p{\frac{q}{1-q}}^{-ue(\Gamma_k)/k}(1-q)^{-\binom{u}{2}}\cdot\frac{|\calG_{n-k,k-u}|}{|\calG_{n,k}|}.\label{eqn:geometric}
\end{align}
Using $\binom{n-k}{k-u}\leq\binom{n}{k}\p{\frac{k}{n-k}}^u$, we get that
\begin{align}
 &\pp{\bE_{\calH_0}(\calN_{\Gamma_k})}^2\cdot\sum_{u=2}^k\binom{k}{u}\p{\frac{q}{1-q}}^{-ue(\Gamma_k)/k}(1-q)^{-\binom{u}{2}}\cdot\frac{|\calG_{n-k,k-u}|}{|\calG_{n,k}|}\nonumber\\
 &\quad\quad\leq \pp{\bE_{\calH_0}(\calN_{\Gamma_k})}^2\cdot\sum_{u=2}^k\binom{k}{u}\p{\frac{q}{1-q}}^{-ue(\Gamma_k)/k}(1-q)^{-\binom{u}{2}}\p{\frac{k}{n-k}}^u\\
 &\quad\quad\leq\pp{\bE_{\calH_0}(\calN_{\Gamma_k})}^2\cdot\sum_{u=2}^k\binom{k}{u}\pp{\p{\frac{q}{1-q}}^{-e(\Gamma_k)/k}(1-q)^{-\frac{u-1}{2}}\p{\frac{k}{n-k}}}^u\\
 & \quad\quad\leq\pp{\bE_{\calH_0}(\calN_{\Gamma_k})}^2\cdot\sum_{u=2}^k\binom{k}{u}\pp{\p{\frac{q}{1-q}}^{-e(\Gamma_k)/k}(1-q)^{-\frac{k-1}{2}}\p{\frac{k}{n-k}}|\s{Aut}(\Gamma_k)|^{1/k}}^u\\
 & \quad\quad= \pp{\bE_{\calH_0}(\calN_{\Gamma_k})}^2\pp{\p{1+\p{\frac{q}{1-q}}^{-e(\Gamma_k)/k}(1-q)^{-\frac{k-1}{2}}\p{\frac{k}{n-k}}|\s{Aut}(\Gamma_k)|^{1/k}}^k-1}\\
 &\quad\quad\leq \pp{\bE_{\calH_0}(\calN_{\Gamma_k})}^2\pp{\exp\pp{\p{\frac{k^2}{n-k}}\p{\frac{q}{1-q}}^{-e(\Gamma_k)/k}(1-q)^{-\frac{k-1}{2}}|\s{Aut}(\Gamma_k)|^{1/k}}-1}.
\end{align}
Plugging the last result in \eqref{eqn:termRHS} we obtain
\begin{align}
\bE_{\calH_0}[\s{L}^2(\s{G})] &\leq 1+ \pp{\exp\pp{\p{\frac{k^2}{n-k}}\p{\frac{q}{1-q}}^{-e(\Gamma_k)/k}(1-q)^{-\frac{k-1}{2}}|\s{Aut}(\Gamma_k)|^{1/k}}-1}.
\end{align}
Now, it is clear that any choice of $k$ that satisfies the condition in the statement of the theorem implies that the term in the exponent is $o(1)$, and so, $\bE_{\calH_0}[\s{L}^2(\s{G})] \leq 1+o(1)$, which concludes the proof. 

An alternative technique to evaluate the above is as follows: note that we may rewrite the summation term in \eqref{eqn:geometric} as follows
\begin{align}
&\sum_{u=2}^k\binom{k}{u}\p{\frac{q}{1-q}}^{-ue(\Gamma_k)/k}(1-q)^{-\binom{u}{2}}\cdot\frac{|\calG_{n-k,k-u}|}{|\calG_{n,k}|}\nonumber\\ &\quad\quad\quad\leq \bE_{\s{U}\sim\s{Geom}(n,k,k)}\pp{\p{\frac{q}{1-q}}^{-e(\Gamma_k)/k}(1-q)^{-\frac{k-1}{2}}}^{\s{U}}-\pr_{\s{U}\sim\s{Geom}(n,k,k)}\pp{\s{U}=0}.
\end{align}
It can be shown that $\pr_{\s{U}\sim\s{Geom}(n,k,k)}\pp{\s{U}=0}\to1$, and it is well-known that for any convex function $f$, we have
\begin{align}
\bE_{\s{U}\sim\s{Geom}(n,k,k)}[f(\s{U})]\leq\bE_{\s{B}\sim\s{Binom}(k,k/n)}[f(\s{B})].
\end{align}
Thus,
\begin{align}
&\bE_{\s{U}\sim\s{Geom}(n,k,k)}\pp{\p{\frac{q}{1-q}}^{-e(\Gamma_k)/k}(1-q)^{-\frac{k-1}{2}}}^{\s{U}}\nonumber\\
&\quad\quad\quad\quad\quad\quad\leq \bE_{\s{B}\sim\s{Binom}(k,k/n)}\pp{\p{\frac{q}{1-q}}^{-e(\Gamma_k)/k}(1-q)^{-\frac{k-1}{2}}}^{\s{B}}\nonumber\\
&\quad\quad\quad\quad\quad\quad= \p{1+\frac{k}{n}\p{\p{\frac{q}{1-q}}^{-e(\Gamma_k)/k}(1-q)^{-\frac{k-1}{2}}-1}}^k\nonumber\\
&\quad\quad\quad\quad\quad\quad\leq \exp\pp{\frac{k^2}{n}\p{\p{\frac{q}{1-q}}^{-e(\Gamma_k)/k}(1-q)^{-\frac{k-1}{2}}-1}},
\end{align}
which converges to $1$ under the condition in Corollary~\ref{cor:1}.

We now consider the general case. Recall that from \eqref{eqn:termRHS} we have
\begin{align}
\bE_{\calH_0}[\s{L}^2(\s{G})] &\leq 1+\frac{1}{\pp{\bE_{\calH_0}(\calN_{\Gamma_k})}^2}\sum_{(\ell_1,\ell_2):\;e(\bar{\Gamma}_{\ell_1}\cap\bar{\Gamma}_{\ell_2})>0}\p{\frac{q}{1-q}}^{e(\bar{\Gamma}_{\ell_1}\cup\bar{\Gamma}_{\ell_2})}(1-q)^{2\cdot\binom{k}{2}-\binom{v(\bar{\Gamma}_{\ell_1}\cap\bar{\Gamma}_{\ell_2})}{2}}\\
& = 1+\frac{1}{|\calG_{n,k}|^2}\sum_{(\ell_1,\ell_2):\;e(\bar{\Gamma}_{\ell_1}\cap\bar{\Gamma}_{\ell_2})>0}\p{\frac{q}{1-q}}^{-e(\bar{\Gamma}_{\ell_1}\cap\bar{\Gamma}_{\ell_2})}(1-q)^{-\binom{v(\bar{\Gamma}_{\ell_1}\cap\bar{\Gamma}_{\ell_2})}{2}}\\
& \triangleq 1+\frac{1}{|\calG_{n,k}|^2}\sum_{(\ell_1,\ell_2):\;e(\bar{\Gamma}_{\ell_1}\cap\bar{\Gamma}_{\ell_2})>0}\bar{f}(\bar{\Gamma}_{\ell_1}\cap\bar{\Gamma}_{\ell_2}).\label{eqn:termRHS2}
\end{align}
We may write:
\begin{align}
\bE_{\calH_0}[\s{L}^2(\s{G})] &\leq 1+ \frac{1}{|\calG_{n,k}|^2}\sum_{(\ell_1,\ell_2):\;e(\bar{\Gamma}_{\ell_1}\cap\bar{\Gamma}_{\ell_2})>0}\bar{f}(\bar{\Gamma}_{\ell_1}\cap\bar{\Gamma}_{\ell_2})\\
& = 1+ \frac{1}{|\calG_{n,k}|^2}\sum_{\s{H}\subseteq\Gamma_k:\;e(\s{H})>0}\sum\limits_{\substack{(\ell_1,\ell_2): \\ \bar{\Gamma}_{\ell_1}\cap\bar{\Gamma}_{\ell_2}\simeq\s{H}}}\bar{f}(\bar{\Gamma}_{\ell_1}\cap\bar{\Gamma}_{\ell_2})\\
& = 1+ \frac{1}{|\calG_{n,k}|^2}\sum_{\s{H}\subseteq\Gamma_k:\;e(\s{H})>0}\bar{f}(\s{H})\cdot\abs{(\ell_1,\ell_2)\in\calG_{n,k}^2:\;\bar{\Gamma}_{\ell_1}\cap\bar{\Gamma}_{\ell_2}\simeq\s{H}},
\end{align}
where for each subgraph $\s{H}\subseteq\s{\Gamma_k}$ we count the number of pairs $(\bar{\Gamma}_{\ell_1},\bar{\Gamma}_{\ell_2})$ of copies of $\Gamma_k$ such that their intersection $\bar{\Gamma}_{\ell_1}\cap\bar{\Gamma}_{\ell_2}$ is isomorphic ($\simeq$) to $\s{H}$. We next upper bound this number pairs. Given $\s{H}$, a pair of copies of $\Gamma_k$ with intersection $\s{H}$ is a structure consisting of $2k-v(\s{H})$ vertices. Accordingly, the number of pair should be proportional to $n^{2k-v(\s{H})}$. Precisely, we can upper bound this quantity by
\begin{align}
\abs{(\ell_1,\ell_2)\in\calG_{n,k}^2:\;\bar{\Gamma}_{\ell_1}\cap\bar{\Gamma}_{\ell_2}\simeq\s{H}}\leq |\calG_{n,k}|\cdot|\calG_{n-k,k-v(\s{H})}|,
\end{align}
where the equality follows from the following reasoning: there are $|\calG_{n,k}|$ ways of picking a graph $\bar{\Gamma}_{\ell_1}$ of $k$ vertices from a graph on $n$ vertices. Then, once $\bar{\Gamma}_{\ell_1}$ have been determined, given the $v(\s{H})$ nodes of $\bar{\Gamma}_{\ell_2}$ that will be shared with $\bar{\Gamma}_{\ell_1}$, it remains to pick from $\bar{\Gamma}_{\ell_1}^c$ the remaining $k-v(\s{H})$ nodes of $\bar{\Gamma}_{\ell_2}$, and there are exactly $|\calG_{n-k,k-v(\s{H})}|$ ways of doing that. Thus, 
\begin{align}
\bE_{\calH_0}[\s{L}^2(\s{G})] &\leq 1+ \frac{1}{|\calG_{n,k}|^2}\sum_{\s{H}\subseteq\Gamma_k:\;e(\s{H})>0}\bar{f}(\s{H})\cdot|\calG_{n,k}|\cdot|\calG_{n-k,k-v(\s{H})}|\\
& = 1+\sum_{\s{H}\subseteq\Gamma_k:\;e(\s{H})>0}\p{\frac{q}{1-q}}^{-e(\s{H})}(1-q)^{-\binom{v(\s{H})}{2}}\frac{|\calG_{n-k,k-v(\s{H})}|}{|\calG_{n,k}|}\\
& \leq 1+\sum_{\s{H}\subseteq\Gamma_k:\;e(\s{H})>0}\p{\frac{q}{1-q}}^{-e(\s{H})}(1-q)^{-\binom{v(\s{H})}{2}}\p{\frac{k}{n-k}}^{v(\s{H})}\\
& = 1+\sum_{\s{H}\subseteq\Gamma_k:\;e(\s{H})>0}\pp{\p{\frac{q}{1-q}}^{-e(\s{H})/v(\s{H})}(1-q)^{-\frac{v(\s{H})-1}{2}}\p{\frac{k}{n-k}}}^{v(\s{H})}\\
& \leq 1\nonumber\\
&+\sum_{\s{H}\subseteq\Gamma_k:\;e(\s{H})>0}\pp{\p{\frac{q}{1-q}}^{-e(\s{H})/v(\s{H})}(1-q)^{-\frac{v(\s{H})-1}{2}}\p{\frac{k}{n-k}}|\s{Aut}(\s{H})|^{1/v(\s{H})}}^{v(\s{H})}\\
&\triangleq 1+\sum_{\s{H}\subseteq\Gamma_k:\;e(\s{H})>0}[g(\s{H})]^{v(\s{H})}\\
&\leq 1+\sum_{\s{H}\subseteq\Gamma_k:\;e(\s{H})>0}\pp{\max_{\s{H}\subseteq\Gamma_k:\;e(\s{H})>0}g(\s{H})}^{v(\s{H})}\\
& = 1+\sum_{\ell=2}^k\binom{k}{\ell}\pp{\max_{\s{H}\subseteq\Gamma_k:\;e(\s{H})>0}g(\s{H})}^{\ell}\\
&  = 1+\pp{\p{1+\max_{\s{H}\subseteq\Gamma_k:\;e(\s{H})>0}g(\s{H})}^k-1}\\
&\leq 1+\pp{\exp\pp{k\cdot\max_{\s{H}\subseteq\Gamma_k:\;e(\s{H})>0}g(\s{H})}-1}.
\end{align}
Finally, it is clear that any choice of $k$ that satisfies the condition in the statement of the theorem implies that the term in the exponent is $o(1)$, and so, $\bE_{\calH_0}[\s{L}^2(\s{G})] \leq 1+o(1)$, which concludes the proof of Theorem~\ref{thm:IT_limit}. 

\subsection{Proof of Theorem~\ref{thm:IT_limit_alog}}

We next analyze Algorithm~\ref{algo:optAlg}, by upper bounding its Type-I+II error probability. Let $\hat{\s{H}}$ be the subgraph in $\Gamma_k$, which achieves the minimum value of $\min_{\s{H}\subseteq\Gamma_k:\;e(\s{H})>0}\s{D}_{\s{H}}$. Note that such a subgraph always exists since we do not exclude $\hat{\s{H}}=\s{H}$. Furthermore, let $\omega_{\hat{\s{H}}}(\s{G})$ denote the $\hat{\s{H}}$-number of $\s{G}$, defined as the size (i.e., number of vertices) of the largest $\hat{\s{H}}$ in the graph. We have
\begin{align}
\pr_{\calH_0}(\s{alg}(\s{G})=1) &= \pr_{\calH_0}(\omega_{\hat{\s{H}}}(\s{G})\geq v(\hat{\s{H}}))\\
& = \pr_{\calH_0}(\s{N}_{v(\hat{\s{H}})}(\hat{\s{H}})\geq1),
\end{align}
where $\s{N}_{v(\hat{\s{H}})}(\hat{\s{H}})$ is the number of $\hat{\s{H}}$-graph of size $v(\hat{\s{H}})$ in $\s{G}$. Then, Markov inequality implies that
\begin{align}
\pr_{\calH_0}(\s{N}_{v(\hat{\s{H}})}(\hat{\s{H}})\geq1)&\leq \bE_{\calH_0}\pp{\s{N}_{v(\hat{\s{H}})}(\hat{\s{H}})}\\
& = \binom{n}{v(\hat{\s{H}})}\frac{v(\hat{\s{H}})!}{|\s{Aut}(\hat{\s{H}})|}\p{\frac{q}{1-q}}^{e(\hat{\s{H}})}(1-q)^{\binom{v(\hat{\s{H}})}{2}},
\label{eqn:Type1}
\end{align}
which goes to zero since $\s{J}_{\Gamma_k}\to0$, as stated in the theorem. On the other hand, under the alternative hypothesis $\calH_1$, it is clear that $\s{G}$ contains a $\Gamma_k$-structure of size $k$ with probability one, and therefore must contain an $\hat{\s{H}}$ subgraph of size $v(\hat{\s{H}})$ as well, making $\pr_{\calH_1}(\s{alg}(\s{G})=0)=0$.
\subsection{Proof of Theorem~\ref{thm:IT_limit_alog_recovery}}
We prove that when the condition in Theorem~\ref{thm:IT_limit_alog_recovery} holds, a random graph $\s{G}$ drawn from the distribution $\calG(n,q,k,\Gamma_k)$ has a unique induced subgraph $\Gamma_k$ of size $k$, with high probability. Thus, the estimator that outputs any subgraph $\Gamma_k$ of size $k$ if such a subgraph exists and the empty set otherwise errs with probability converging zero. Let $\s{H}$ be the minimizer of $\min_{\s{H}\subseteq\Gamma_k:\;e(\s{H})>0}\s{D}_{\s{H}}$. Then, it is clear that,
\begin{align}
\pr_{\calH_1}\pp{\s{G}\text{ has two }\Gamma_k\text{-structures of size }k} &\leq \pr_{\calH_1}\pp{\s{G}\text{ has two }\s{H}\text{-structures of size }v(\s{H})}.
\end{align}
By symmetry we may fix the hidden structure $\s{H}_0$ on vertices $\{1,2,\ldots,v(\s{H})\}$. Thus,
\begin{align}
&\pr_{\calH_1}\pp{\s{G}\text{ has two }\s{H}\text{-structures of size }v(\s{H})}\nonumber\\
&\quad\quad\quad\quad= \pr_{\s{H}_0}\pp{\s{G}\text{ has a }\s{H}\text{-structure of size }v(\s{H})\text{ different from }\s{H}_0}\\
&\quad\quad\quad\quad\leq \bE_{\s{H}_0}[\calN_{v(\s{H})}^{\neq\s{H}_0}],
\end{align}
where the inequality follows from Markov's inequality, $\bE_{\s{H}_0}$ denotes expectation with respect to the distribution of $\s{G}$ with $\s{H}_0$ being planted inside, and $\calN_{v(\s{H})}^{\neq\s{H}_0}$ denotes the number of $\s{H}$-structures of size $v(\s{H})$ in $\s{G}$ that are different from $\s{H}_0$. A simple counting argument implies that
\begin{align}
\bE_{\s{H}_0}[\calN_{v(\s{H})}^{\neq\s{H}_0}] &= \sum_{\ell=2}^{|\calG_{n,v(\s{H})}|}\pr_{\s{H}_0}\pp{\s{H}_{\ell}\in\s{G}} \\
& = \sum_{\ell=2}^{|\calG_{n,v(\s{H})}|}q^{e(\s{H}_\ell\setminus\s{H}_0)}(1-q)^{e(\s{H}_\ell^c\setminus\s{H}_0^c)}\\
& = \p{\frac{q}{1-q}}^{e(\s{H})}(1-q)^{\binom{v(\s{H})}{2}}\sum_{\ell=2}^{|\calG_{n,v(\s{H})}|}\p{\frac{q}{1-q}}^{-e(\s{H}_\ell\cap\s{H}_0)}(1-q)^{-\binom{v(\s{H}_\ell\cap\s{H}_0)}{2}}\\
& \triangleq f(\s{H})\cdot \sum_{\ell=2}^{|\calG_{n,v(\s{H})}|}f^{-1}(\s{H}_\ell\cap\s{H}_0).
\end{align}
It suffices to show that $\bE_{\s{H}_0}[\calN_{v(\s{H})}^{\neq\s{H}_0}]\to0$ whenever $\s{v}(\s{H})$ satisfies $\s{J}_{\Gamma_k}\to0$. Summing over all possible isomorphism types of intersections, we get
\begin{align}
\bE_{\s{H}_0}[\calN_{v(\s{H})}^{\neq\s{H}_0}] &= f(\s{H})\sum_{\s{H}'\subseteq\s{H}:\;e(\s{H}')<e(\s{H})}f^{-1}(\s{H}')\cdot\abs{\ell\in\calG_{n,v(\s{H})}:\;\s{H}_{\ell}\cap\s{H}_{0}\simeq\s{H}'}\\
& \leq  f(\s{H})\sum_{\s{H}'\subseteq\s{H}:\;e(\s{H}')<e(\s{H})}f^{-1}(\s{H}')\cdot|\calG_{n-v(\s{H}),v(\s{H})-v(\s{H}')}|.
\end{align}
First we examine the case when $v(\s{H})$ is large, say $v(\s{H})\geq v_0$, where $v_0$ is to be determined in the sequel. In this case, we have
\begin{align}
\bE_{\s{H}_0}[\calN_{v(\s{H})}^{\neq\s{H}_0}]&\leq f(\s{H})\sum_{\s{H}'\subseteq\s{H}:\;e(\s{H}')<e(\s{H})}f^{-1}(\s{H}')\cdot|\calG_{n-v(\s{H}),v(\s{H})-v(\s{H}')}|\\
& = \sum_{\s{H}'\subseteq\s{H}:\;e(\s{H}')<e(\s{H})}\p{\frac{q}{1-q}}^{e(\s{H})-e(\s{H}')}(1-q)^{\binom{v(\s{H})}{2}-\binom{v(\s{H}')}{2}}\cdot|\calG_{n-v(\s{H}),v(\s{H})-v(\s{H}')}|\\
&\leq \sum_{\s{H}'\subseteq\s{H}:\;e(\s{H}')<e(\s{H})}(1-q)^{\binom{v(\s{H})}{2}-\binom{v(\s{H}')}{2}}n^{v(\s{H})-v(\s{H}')}\\
&=\sum_{m=0}^{v(\s{H})-1}\binom{v(\s{H})}{m}(1-q)^{\binom{v(\s{H})}{2}-\binom{m}{2}}n^{v(\s{H})-m}\\
&\leq \sum_{m=0}^{v(\s{H})-1}\pp{n\cdot v(\s{H})\cdot(1-q)^{\frac{v(\s{H})+v(\s{H}')-1}{2}}}^{v(\s{H})-m}\\
& \leq \sum_{m=1}^{v(\s{H})}\pp{n\cdot v(\s{H})\cdot(1-q)^{\frac{v(\s{H})+1}{2}}}^{m}\\
&\leq \frac{1}{1-n\cdot v(\s{H})\cdot(1-q)^{\frac{v(\s{H})+1}{2}}}-1\label{eqn:goesto0}.
\end{align}
We next choose $v_0$ such that the r.h.s. of \eqref{eqn:goesto0} converges to zero. To that end, it is easy to show that it is sufficient to take $v_0\geq 6\cdot\log_{1/(1-q)} n$. Suppose now that $v(\s{H})<v_0$. For some $\ell\in\mathbb{N}$ to be defined below, we split the sum as follows:
\begin{align}
\bE_{\s{H}_0}[\calN_{v(\s{H})}^{\neq\s{H}_0}] &= f(\s{H})\sum_{\s{H}'\subseteq\s{H}:\;e(\s{H}')<e(\s{H}),v(\s{H}')\leq\ell}f^{-1}(\s{H}')\cdot|\calG_{n-v(\s{H}),v(\s{H})-v(\s{H}')}|\nonumber\\
&\quad+f(\s{H})\sum_{\s{H}'\subseteq\s{H}:\;e(\s{H}')<e(\s{H}),v(\s{H}')>\ell}f^{-1}(\s{H}')\cdot|\calG_{n-v(\s{H}),v(\s{H})-v(\s{H}')}|.\label{eqn:twoterm}
\end{align}
Now, for the first term, we have
\begin{align}
&f(\s{H})\sum_{\s{H}'\subseteq\s{H}:\;e(\s{H}')<e(\s{H}),v(\s{H}')\leq\ell}f^{-1}(\s{H}')\cdot|\calG_{n-v(\s{H}),v(\s{H})-v(\s{H}')}| \nonumber\\
&= f(\s{H})\cdot|\calG_{n,v(\s{H})}|\cdot\sum_{\s{H}'\subseteq\s{H}:\;e(\s{H}')<e(\s{H}),v(\s{H}')\leq\ell}f^{-1}(\s{H}')\cdot\frac{|\calG_{n-v(\s{H}),v(\s{H})-v(\s{H}')}|}{|\calG_{n,v(\s{}H)}|}\\
&\leq f(\s{H})\cdot|\calG_{n,v(\s{H})}|\cdot\sum_{\s{H}'\subseteq\s{H}:\;e(\s{H}')<e(\s{H}),v(\s{H}')\leq\ell}f^{-1}(\s{H}')\cdot\p{\frac{v(\s{H})}{n-v(\s{H})}}^{v(\s{H}')}\\
& = f(\s{H})\cdot|\calG_{n,v(\s{H})}|\nonumber\\
&\quad\quad\cdot\sum_{\s{H}'\subseteq\s{H}:\;e(\s{H}')<e(\s{H}),v(\s{H}')\leq\ell}\pp{\frac{v(\s{H})}{n-v(\s{H})}\p{\frac{q}{1-q}}^{-e(\s{H}')/v(\s{H}')}(1-q)^{-\frac{v(\s{H}')-1}{2}}}^{v(\s{H}')}\\
&\leq f(\s{H})\cdot|\calG_{n,v(\s{H})}|\cdot\sum_{\s{H}'\subseteq\s{H}:\;e(\s{H}')<e(\s{H}),v(\s{H}')\leq\ell}\pp{\frac{v(\s{H})}{n-v(\s{H})}q^{-\frac{v(\s{H}')-1}{2}}}^{v(\s{H}')}\\
& \leq f(\s{H})\cdot|\calG_{n,v(\s{H})}|\cdot\sum_{m=0}^{\ell}\pp{\frac{v(\s{H})^2}{n-v(\s{H})}q^{-\frac{v(\s{H}')-1}{2}}}^{m}\\
&\leq f(\s{H})\cdot|\calG_{n,v(\s{H})}|\cdot\sum_{m=0}^{\ell}\pp{\frac{v(\s{H})^2}{n-v(\s{H})}q^{-\frac{\ell-1}{2}}}^{m},\label{eqn:partialSumrecovery}
\end{align}
where in the second inequality we have used the fact that $e(\s{H}')\leq\binom{v(\s{H}')}{2}$. From the theorem statement we know that $f(\s{H})\cdot|\calG_{n,v(\s{H})}|\to0$, and as so in order to show that the r.h.s. of \eqref{eqn:partialSumrecovery} is converging to zero, it is sufficient to show that the summation term at the r.h.s. of \eqref{eqn:partialSumrecovery} is bounded. However, a simple calculation shows that by taking any $\ell\leq 2\cdot\log_{1/q}n-2\log_{1/q}[6\log_{1/(1-q)}n]+1$, we have $q^{-\frac{\ell-1}{2}}\leq \frac{n-v(\s{H})}{v(\s{H})^2}$, which implies that the summation term at the r.h.s. of \eqref{eqn:partialSumrecovery} is indeed bounded. It remains to bound second term on the r.h.s. of \eqref{eqn:twoterm}. We may write
\begin{align}
&f(\s{H})\sum_{\s{H}'\subseteq\s{H}:\;e(\s{H}')<e(\s{H}),v(\s{H}')>\ell}f^{-1}(\s{H}')\cdot|\calG_{n-v(\s{H}),v(\s{H})-v(\s{H}')}| \nonumber\\
& = \sum_{\s{H}'\subseteq\s{H}:\;e(\s{H}')<e(\s{H}),v(\s{H}')>\ell}\p{\frac{q}{1-q}}^{e(\s{H})-e(\s{H}')}(1-q)^{\binom{v(\s{H})}{2}-\binom{v(\s{H}')}{2}}\cdot|\calG_{n-v(\s{H}),v(\s{H})-v(\s{H}')}|\\
& \leq \sum_{\s{H}'\subseteq\s{H}:\;e(\s{H}')<e(\s{H}),v(\s{H}')>\ell}(1-q)^{\binom{v(\s{H})}{2}-\binom{v(\s{H}')}{2}}\cdot|\calG_{n-v(\s{H}),v(\s{H})-v(\s{H}')}|\\
& =\sum_{m=\ell+1}^{v(\s{H})-1}\binom{v(\s{H})}{\ell}(1-q)^{\binom{v(\s{H})}{2}-\binom{m}{2}}\cdot|\calG_{n-v(\s{H}),v(\s{H})-m}|\\
& \leq\sum_{m=\ell+1}^{v(\s{H})-1}\frac{v(\s{H})^{v(\s{H})-m}n^{v(\s{H})-m}}{(v(\s{H})-m)!^2}(1-q)^{\binom{v(\s{H})}{2}-\binom{m}{2}}\\
& \leq \sum_{m=\ell+1}^{v(\s{H})-1}\pp{\frac{n\cdot v(\s{H})e^2}{(v(\s{H})-m)^2}\cdot(1-q)^{\frac{v(\s{H})+m-1}{2}}}^{v(\s{H})-m}.
\end{align}
To finish the proof, it suffices to show that the expression within the parentheses goes to zero uniformly for all $m>\ell$. But this follows since for $m\geq v(\s{H})/2$,
\begin{align}
\p{\frac{1}{1-q}}^{\frac{v(\s{H})+m-1}{2}}\geq \p{\frac{1}{1-q}}^{3v(\s{H})/2-1}\gg n\cdot v(\s{H}),
\end{align}
and for $m\in(\ell,v(\s{H})/2)$,
\begin{align}
\p{\frac{1}{1-q}}^{\frac{v(\s{H})+m-1}{2}}\geq \p{\frac{1}{1-q}}^{v(\s{H})/2+\ell/2-1}\gg \frac{n\cdot v(\s{H})}{(v(\s{H})-m)^2}.
\end{align}

\subsection{Recovery Lower Bound}

Let us define the following quantity:
\begin{align}
\s{Int}_{\s{G}}(\Gamma_k)\triangleq \frac{1}{\calN^2_{\Gamma_k}}\sum_{\ell_1=1}^{|\calG_{n,k}|}\sum_{\ell_1=2}^{|\calG_{n,k}|}\Ind\pp{\bar{\Gamma}_{\ell_1}\cap\bar{\Gamma}_{\ell_2}\neq\emptyset},
\end{align}
where $\bar{\Gamma}_1,\bar{\Gamma}_2,\ldots,\bar{\Gamma}_{|\calG_{n,k}|}$ are all possible subgraph copies of $\Gamma_k$ in the complete graph, and $\calN_{\Gamma_k}\triangleq \sum_{\ell=1}^{|\calG_{n,k}|}\Ind\pp{\bar{\Gamma}_\ell\in\s{G}}$. Namely, $\s{Int}_{\s{G}}(\Gamma_k)$ is the proportion of pairs copies of $\Gamma_k$ in $\s{G}$ whose intersection is nonempty. Let $\s{N}_\ell\triangleq\Ind\pp{\bar{\Gamma}_\ell\in\s{G}}$. Then,
\begin{align}
\bE_{\calH_0}[\calN^2_{\Gamma_k}] &= \sum_{\ell_1,\ell_2}\pr_{\calH_0}[\s{N}_{\ell_1}\s{N}_{\ell_2}] \\
&= \sum_{\ell_1,\ell_2:\bar{\Gamma}_{\ell_1}\cap\bar{\Gamma}_{\ell_2}=\emptyset}\pr_{\calH_0}[\s{N}_{\ell_1}\s{N}_{\ell_2}]+\sum_{\ell_1,\ell_2:\bar{\Gamma}_{\ell_1}\cap\bar{\Gamma}_{\ell_2}\neq\emptyset}\pr_{\calH_0}[\s{N}_{\ell_1}\s{N}_{\ell_2}]\\
& \triangleq \s{A}+\s{B}.
\end{align}
We can easily compute $\s{A}$. Indeed,
\begin{align}
\s{A} = |\calG_{n,k}|\cdot|\calG_{n-k,k}|\p{\frac{q}{1-q}}^{2e(\Gamma_k)}(1-q)^{2\binom{k}{2}}\sim\pp{\bE_{\calH_0}(\calN_{\Gamma_k})}^2.
\end{align}
Now, recall that the likelihood function is defined as $\s{L}(\s{G}) = \calN_{\Gamma_k}/\bE_{\calH_0}(\calN_{\Gamma_k})$, and that under the conditions of Theorem~\ref{thm:IT_limit}, we have $\bE_{\calH_0}[\s{L}(\s{G})]^2=1+o(1)$. Therefore, it follows that
\begin{align}
\frac{\s{B}}{\pp{\bE_{\calH_0}(\calN_{\Gamma_k})}^2}=o(1).\label{eqn:Basymp}
\end{align}
Next, we note that
\begin{align}
\s{Int}_{\s{G}}(\Gamma_k)= \frac{1}{\calN^2_{\Gamma_k}}\sum_{\ell_1=1}^{|\calG_{n,k}|}\sum_{\ell_1=2}^{|\calG_{n,k}|}\Ind\pp{\bar{\Gamma}_{\ell_1}\cap\bar{\Gamma}_{\ell_2}\neq\emptyset} = \frac{1}{\calN^2_{\Gamma_k}}\sum_{\ell_1,\ell_2:\bar{\Gamma}_{\ell_1}\cap\bar{\Gamma}_{\ell_2}\neq\emptyset}\s{N}_{\ell_1}\s{N}_{\ell_2}.
\end{align}
Therefore, we can decompose $\s{Int}_{\s{G}}(\Gamma_k)$ as follows,
\begin{align}
\s{Int}_{\s{G}}(\Gamma_k) &= \s{Int}_{\s{G}}(\Gamma_k)\Ind\pp{\s{L}^2(\s{G})\geq1/2}+ \s{Int}_{\s{G}}(\Gamma_k)\Ind\pp{\s{L}^2(\s{G})<1/2}\\
& = \frac{\sum_{\ell_1,\ell_2:\bar{\Gamma}_{\ell_1}\cap\bar{\Gamma}_{\ell_2}\neq\emptyset}\s{N}_{\ell_1}\s{N}_{\ell_2}}{\bE[\calN^2_{\Gamma_k}]}\frac{1}{\s{L^2(G)}}\Ind\pp{\s{L}^2(\s{G})\geq1/2}+ \s{Int}_{\s{G}}(\Gamma_k)\Ind\pp{\s{L}^2(\s{G})<1/2},
\end{align}
and thus,
\begin{align}
\bE_{\calH_0}[\s{Int}_{\s{G}}(\Gamma_k)]&\leq \bE_{\calH_0}\pp{\frac{\sum_{\ell_1,\ell_2:\bar{\Gamma}_{\ell_1}\cap\bar{\Gamma}_{\ell_2}\neq\emptyset}\s{N}_{\ell_1}\s{N}_{\ell_2}}{\bE[\calN^2_{\Gamma_k}]}\frac{1}{\s{L^2(G)}}\Ind\pp{\s{L}^2(\s{G})\geq1/2}}\nonumber\\
&\quad+\bE_{\calH_0}\pp{\s{Int}_{\s{G}}(\Gamma_k)\Ind\pp{\s{L}^2(\s{G})<1/2}}\\
&\leq 2\cdot\bE_{\calH_0}\pp{\frac{\sum_{\ell_1,\ell_2:\bar{\Gamma}_{\ell_1}\cap\bar{\Gamma}_{\ell_2}\neq\emptyset}\s{N}_{\ell_1}\s{N}_{\ell_2}}{\bE[\calN^2_{\Gamma_k}]}}+\bE_{\calH_0}\pp{\Ind\pp{\s{L}^2(\s{G})<1/2}}\\
&= 2\cdot \frac{\s{B}}{\pp{\bE_{\calH_0}(\calN_{\Gamma_k})}^2}+\pr_{\calH_0}\p{\s{L}^2(\s{G})<1/2}\\
& \leq o(1),\label{eqn:ExpectedOverlap}
\end{align}
where in the inequality we have used the fact that $\s{Int}_{\s{G}}(\Gamma_k)\leq1$ with probability 1, and the last equality is due to \eqref{eqn:Basymp} and Chebyshev's inequality. In order to prove that exact resconstruction is impossible we will look the following overlap measure,
\begin{align}
\s{over}(\hat{\Gamma}_k)\triangleq \sum_{i\in[n]}\pr_{\calH_1}(i\in\Gamma_k\cap\hat{\Gamma}_k),
\end{align}
where $\hat{\Gamma}_k$ is any possible estimator of $\Gamma_k$. We will show that $\s{over}(\hat{\Gamma}_k)=o(k)$, which implies that exact reconstruction is impossible. To that end, we note that $\s{over}(\hat{\Gamma}_k)$ can be rewritten as follows
\begin{align}
\s{over}(\hat{\Gamma}_k) &= \sum_{\s{G}}\pr_{\calH_1}(\s{G})\sum_{\ell=1}^{|\calG_{n,k}|}\pr_{\calH_1}(\bar{\Gamma}_\ell\vert\s{G})|\bar{\Gamma}_\ell\cap\hat{\Gamma}_k|\\
& = \sum_{\s{G}}\pr_{\calH_1}(\s{G})\sum_{\ell=1}^{|\calG_{n,k}|}\frac{|\bar{\Gamma}_\ell\cap\hat{\Gamma}_k|}{\calN_{\Gamma_k}}\\
& = \sum_{\s{G}}\pr_{\calH_0}(\s{G})\sum_{\ell=1}^{|\calG_{n,k}|}\frac{|\bar{\Gamma}_\ell\cap\hat{\Gamma}_k|}{\calN_{\Gamma_k}}+\sum_{\s{G}}[\pr_{\calH_1}(\s{G})-\pr_{\calH_0}(\s{G})]\sum_{\ell=1}^{|\calG_{n,k}|}\frac{|\bar{\Gamma}_\ell\cap\hat{\Gamma}_k|}{\calN_{\Gamma_k}}\\
&\leq \sum_{\s{G}}\pr_{\calH_0}(\s{G})\sum_{\ell=1}^{|\calG_{n,k}|}\frac{|\bar{\Gamma}_\ell\cap\hat{\Gamma}_k|}{\calN_{\Gamma_k}}+k\cdot\s{TV}(\pr_{\calH_0},\pr_{\calH_1}),
\end{align}
where in the last inequality we have used the definition of the total-variation distance, and the fact that $|\bar{\Gamma}_\ell\cap\hat{\Gamma}_k|\leq k$, for any $\bar{\Gamma}_\ell$ and $\hat{\Gamma}_k$. Since $\s{TV}(\pr_{\calH_0},\pr_{\calH_1})\leq\sqrt{\bE_{\calH_0}(\s{L}^2(\s{G}))-1}$, the condition $\s{J}_{\Gamma_k}\to\infty$ and the proof of Theorem~\ref{thm:IT_limit} imply that $\s{TV}(\pr_{\calH_0},\pr_{\calH_1}) = o(1)$, and therefore,
\begin{align}
\s{over}(\hat{\Gamma}_k)\leq \sum_{\s{G}}\pr_{\calH_0}(\s{G})\sum_{\ell=1}^{|\calG_{n,k}|}\frac{|\bar{\Gamma}_\ell\cap\hat{\Gamma}_k|}{\calN_{\Gamma_k}}+o(k).
\end{align}
Next, we can write
\begin{align}
\s{over}(\hat{\Gamma}_k)&\leq \bE_{\calH_0}\pp{\sum_{\ell=1}^{|\calG_{n,k}|}\frac{|\bar{\Gamma}_\ell\cap\hat{\Gamma}_k|}{\calN_{\Gamma_k}}}+o(k)\\
& = \sum_{i=1}^n\bE_{\calH_0}\pp{\Ind\pp{i\in\hat{\Gamma}_k}\sum_{\ell=1}^{|\calG_{n,k}|}\frac{\Ind\pp{i\in\bar{\Gamma}_\ell}}{\calN_{\Gamma_k}}}+o(k),
\end{align}
and we note that
\begin{align}
\pp{\sum_{\ell=1}^{|\calG_{n,k}|}\frac{\Ind\pp{i\in\bar{\Gamma}_\ell}}{\calN_{\Gamma_k}}}^2 &= \sum_{\ell_1=1}^{|\calG_{n,k}|}\sum_{\ell_2=1}^{|\calG_{n,k}|}\frac{\Ind\pp{i\in\bar{\Gamma}_{\ell_1}}\Ind\pp{i\in\bar{\Gamma}_{\ell_2}}}{\calN^2_{\Gamma_k}}\\
&\leq \s{Int}_{\s{G}}(\Gamma_k).
\end{align}
Thus,
\begin{align}
\s{over}(\hat{\Gamma}_k)&\leq    \sum_{i=1}^n\bE_{\calH_0}\pp{\Ind\pp{i\in\hat{\Gamma}_k}\sqrt{\s{Int}_{\s{G}}(\Gamma_k)}}+o(k)\\
&\leq k\cdot \bE_{\calH_0}\pp{\sqrt{\s{Int}_{\s{G}}(\Gamma_k)}}+o(k)\\
&\leq k\cdot\sqrt{\bE_{\calH_0}\pp{\s{Int}_{\s{G}}(\Gamma_k)}}+o(k)\\
&\leq o(k),
\end{align}
where the third inequality follows from Jensen's inequality, and the last inequality is due to \eqref{eqn:ExpectedOverlap}.

\subsection{Proof of Theorem~\ref{thm:4}}

We start by assuming that $q<1/2$ and $e(\Gamma_k)-q\cdot\binom{k}{2}\geq0$. The complement case, $q<1/2$ and $e(\Gamma_k)-q\cdot\binom{k}{2}\leq0$ is handled in the same way. Recall that under the null hypothesis, the total number of edges $\s{W}(\s{G})$ in $\s{G}$ is distributed as $\s{W}(\s{G})\sim\s{Binomial}\p{\binom{n}{2},q}$, while under the alternative hypothesis $\s{W}(\s{G})\sim e(\Gamma_k)+\s{Binomial}\p{\binom{n}{2}-\binom{k}{2},q}$. The test we would like to analyze is 
\begin{align}
\phi_{\s{Tot}}(\s{G})=1\quad\s{iff}\quad\s{W}(\s{G})\geq\s{W}^\star = q\cdot\binom{n}{2}+\frac{e(\Gamma_k)-q\cdot\binom{k}{2}}{2}.
\end{align}
The average risk is $\gamma_n(\phi_{\s{Tot}})= \pr_{\calH_0}[\phi_{\s{Tot}}(\s{G})=1]+\pr_{\calH_1}[\phi_{\s{Tot}}(\s{G})=0]$. p). By Bernstein's inequality, we have 
\begin{align}
\pr_{\calH_0}[\phi_{\s{Tot}}(\s{G})=1] &= \pr_{\calH_0}\pp{\s{W}(\s{G})\geq\s{W}^\star}\\
& = \pr\pp{\s{Binomial}\p{\binom{n}{2},q}-q\binom{n}{2}\geq \s{W}^\star-q\binom{n}{2}}\\
&\leq \exp\p{-\frac{\pp{\s{W}^\star-q\binom{n}{2}}^2/4}{2\binom{n}{2}q(1-q)+\frac{1}{3}\pp{\s{W}^\star-q\binom{n}{2}}}}\\
& = \exp\p{-\frac{\pp{e(\Gamma_k)-q\cdot\binom{k}{2}}^2}{8\binom{n}{2}q(1-q)+\frac{4}{3}\pp{e(\Gamma_k)-q\cdot\binom{k}{2}}}}.
\end{align}
On the other hand, by the multiplicative Chernoff's bound, it follows that
\begin{align}
\pr_{\calH_1}[\phi_{\s{Tot}}(\s{G})=0] &= \pr_{\calH_1}\pp{\s{W}(\s{G})\leq\s{W}^\star}\\
& = \pr\pp{e(\Gamma_k)+\s{Binomial}\p{\binom{n}{2}-\binom{k}{2},q}\leq \s{W}^\star}\\
&\leq \exp\p{-\frac{\pp{e(\Gamma_k)-q\cdot\binom{k}{2}}^2/4}{2\binom{n}{2}q+2\pp{e(\Gamma_k)-q\cdot\binom{k}{2}}}}.
\end{align}
Therefore, the total average risk is bounded by
\begin{align}
\gamma_n(\phi_{\s{Tot}})\leq 2\cdot\exp\p{-\frac{1}{8}\frac{\pp{e(\Gamma_k)-q\cdot\binom{k}{2}}^2}{q\cdot\binom{n}{2}+\pp{e(\Gamma_k)-q\cdot\binom{k}{2}}}}. 
\end{align}
Accordingly, it is clear that $\gamma_n(\phi_{\s{Tot}})\leq\delta$, for any $\delta\in(0,1)$, if \eqref{eqn:totalfinite} holds.

Next, we consider the case where $q>1/2$ and $e(\Gamma_k^c)-(1-q)\cdot\binom{k}{2}\geq0$. The complement case, $q>1/2$ and $e(\Gamma_k^c)-(1-q)\cdot\binom{k}{2}\leq0$ is handled in the same way. Recall that under the null hypothesis, the total number of edges $\s{W}(\s{G}^c)$ in $\s{G}^c$ is distributed as $\s{W}(\s{G}^c)\sim\s{Binomial}\p{\binom{n}{2},1-q}$, while under the alternative hypothesis $\s{W}(\s{G}^c)\sim e(\Gamma_k^c)+\s{Binomial}\p{\binom{n}{2}-\binom{k}{2},1-q}$. The test we would like to analyze is 
\begin{align}
\phi_{\s{Tot}}(\s{G})=1\quad\s{iff}\quad\s{W}(\s{G})\geq\overline{\s{W}}^\star = (1-q)\cdot\binom{n}{2}+\frac{e(\Gamma_k^c)-(1-q)\cdot\binom{k}{2}}{2}.
\end{align}
Following the same analysis as above we obtain that 
\begin{align}
\gamma_n(\phi_{\s{Tot}})\leq 2\cdot\exp\p{-\frac{1}{8}\frac{\pp{e(\Gamma_k)-q\cdot\binom{k}{2}}^2}{(1-q)\cdot\binom{n}{2}+\pp{e(\Gamma_k)-q\cdot\binom{k}{2}}}},     
\end{align}
where we have used the fact that $e(\Gamma_k^c)-(1-q)\cdot\binom{k}{2} = q\binom{k}{2}-e(\Gamma_k)$. This concludes the proof.

\subsection{Proof of Theorem~\ref{thm:5}}

We start by repeating the same arguments discussed right before the statement of Theorem~\ref{thm:5}. Specifically, under $\calH_1$ consider the unit vector $\mathbf{x}_{\Gamma_k}$ with entries $(x_1,\ldots,x_n)$ such that $x_i=1/\sqrt{k}$ if $i\in v(\Gamma_k)$, and $x_i=0$, otherwise. Then,
\begin{align}
\s{S}(\mathbf{A}) &= \norm{\mathbf{A}-\bE_{\calH_0}\mathbf{A}}_{\s{op}}+\norm{\mathbf{A}^c-\bE_{\calH_0}\mathbf{A}^c}_{\s{op}}\\
&\geq |\mathbf{x}_{\Gamma_k}^T(\mathbf{A}-\bE_{\calH_0}\mathbf{A})\mathbf{x}_{\Gamma_k}|+|\mathbf{x}_{\Gamma_k}^T(\mathbf{A}^c-\bE_{\calH_0}\mathbf{A}^c)\mathbf{x}_{\Gamma_k}| \\
& = \abs{\frac{2e(\Gamma_k)}{k} - (k-1)q} + \abs{\frac{2e(\Gamma_k^c)}{k} - (k-1)(1-q)}\\
& = 2\abs{\frac{2e(\Gamma_k)}{k} - (k-1)q}.\label{eqn:spectlower2proof}
\end{align}
Thus, under the alternative hypothesis, $\pr_{\calH_1}[\s{S}(\mathbf{A})\geq 2\abs{2e(\Gamma_k)/ - (k-1)q}]=1$. 

Next, we examine $\s{S}(\mathbf{A})=\norm{\mathbf{A}-\bE_{\calH_0}\mathbf{A}}_{\s{op}}+\norm{\mathbf{A}^c-\bE_{\calH_0}\mathbf{A}^c}_{\s{op}}$ under the null hypothesis. Note that each of the random matrices inside the spectral norms are symmetric, have zero-mean, and bounded entries. We recall the following well-known concentration result \cite{Furedi81theeigenvalues,vu2005spectral}.\footnote{It should be emphasized that, Lemma~\ref{lem:Kulmus} as stated here, is slightly different from \cite[Theorem 2]{Furedi81theeigenvalues}. Nonetheless, Lemma~\ref{lem:Kulmus} can be derived easily from the proof of \cite[Theorem 2]{Furedi81theeigenvalues}. Specifically, to obtain the result given here, we extract the value of ``$v$" in \cite[Subsection 3.3]{Furedi81theeigenvalues} satisfying $\pr(\max|\lambda|>2\sigma\sqrt{n}+v)\leq\delta$.}
\begin{lemma}\label{lem:Kulmus}
Let $\mathbf{B} = [\mathbf{B}_{ij}]\in\mathbb{R}^{n\times n}$ be a random symmetric matrix where $\mathbf{B}_{ij}$ are independent random variables, for $1\leq i\leq j\leq n$. Assume that there exists $\s{K},\sigma>0$ such that the following conditions hold
\begin{enumerate}
\item $\bE \mathbf{B}_{ij} = 0$, for $1\leq i\leq j\leq n$.
\item $|\mathbf{B}_{ij}| \leq \s{K}$, for $1\leq i\leq j\leq n$.
\item $\bE\mathbf{B}_{ij}^2\leq\sigma^2$, for $1\leq i\leq j\leq n$.
\end{enumerate}
Then, for any $\delta>0$,
\begin{align}
\norm{\mathbf{B}}_{\s{opt}}\leq 2\sigma\sqrt{n}+\frac{\sigma\sqrt{n}\log\frac{n}{\delta^2}}{\p{\frac{\sigma}{\s{K}}}^{1/3}n^{1/6}-\frac{1}{2}\log\frac{n}{\delta^2}},
\end{align}
with probability at least $1-\delta$.
\end{lemma}

Our matrices $\mathbf{A}-\bE_{\calH_0}\mathbf{A}$ and $\mathbf{A}^c-\bE_{\calH_0}\mathbf{A}^c$ satisfy the conditions of Lemma~\ref{lem:Kulmus}: 1) both matrices have zero mean, 2) for both matrices $\s{K}=1$, and finally 3) for both matrices we clearly have $\sigma = \sqrt{q(1-q)}$. Therefore, Lemma~\ref{lem:Kulmus} implies that with probability at least $1-\delta/2$,
\begin{align}
\norm{\mathbf{A}-\bE_{\calH_0}\mathbf{A}}_{\s{op}}\leq 2\cdot\sqrt{q(1-q)n}+\frac{\sqrt{q(1-q)n}\log\frac{4n}{\delta^2}}{[q(1-q)n]^{1/6}-\frac{1}{2}\log\frac{4n}{\delta^2}},
\end{align}
and the same upper bound holds for $\norm{\mathbf{A}^c-\bE_{\calH_0}\mathbf{A}^c}_{\s{op}}$, for any $\delta>0$. Therefore,
with probability at least $1-\delta$,
\begin{align}
\s{S}(\mathbf{A})\leq 4\cdot\sqrt{q(1-q)n}+2\frac{\sqrt{q(1-q)n}\log\frac{4n}{\delta^2}}{[q(1-q)n]^{1/6}-\frac{1}{2}\log\frac{4n}{\delta^2}}\triangleq \varphi(n,q,\delta).\label{eqn:normUpperBound}
\end{align}
Thus, using \eqref{eqn:spectlower2proof} and \eqref{eqn:normUpperBound} we may conclude that the spectral test $\phi_{\s{spec}}$ that accepts the null hypothesis
iff $\s{S}(\mathbf{A})\leq\varphi(n,q,\delta)$, achieves average risk $\gamma_n (\phi_{\s{spec}})\leq\delta$, if
\begin{align}
2\abs{\frac{2e(\Gamma_k)}{k} - (k-1)q}\geq \varphi(n,q,\delta),
\end{align}
as stated in Theorem~\ref{thm:5}.

\subsection{Proof of Theorem~\ref{thm:6}}

Our proof of Theorem~\ref{thm:6} will follow the strategy of, e.g., \cite{Hopkins18}, of expanding the likelihood ration $\s{L}_n$ in a basis of orthogonal polynomials with respect to $\pr_{\calH_0}$. Specifically, suppose that $f_0,f_1,\ldots,f_m:\Omega^n\to\mathbb{R}$ are an orthonormal basis for the coordinate-degree $\s{D}$ functions (with respect to $\left\langle \cdot,\cdot \right\rangle_{\calH_0}$), and that $f_0$ is the unit constant function. Therefore, $\left\langle f_i,f_j \right\rangle_{\calH_0}=\delta_{ij}$, where $\delta_{ij}=0$ if $i\neq j$, and $\delta_{ii}=1$. Then, measuring the norm of $\s{L}_{n,\leq\s{D}}$ in this basis, we have
\begin{align}
\norm{\s{L}_{n,\leq\s{D}}}_{\calH_0}^2 &= \sum_{1\leq i\leq m}\left\langle f_i,\s{L}_{n,\leq\s{D}} \right\rangle_{\calH_0}^2\\
& = \sum_{1\leq i\leq m}\pp{\bE_{\calH_0}\pp{\s{L}_n(\s{G})f_i(\s{G})}}^2\\
& = \sum_{1\leq i\leq m}\pp{\bE_{\calH_1}f_i(\s{G})}^2,\label{eqn:orthogonalDecomposition}
\end{align}
where we have used the fact that $(\s{L}_n-\s{L}_{n,\leq\s{D}})$ is orthogonal to $\{f_i\}_{i=0}^m$. Therefore, to prove Theorem~\ref{thm:6} we only need to compute $\bE_{\calH_1}f_i(\s{G})$ for some orthonormal basis functions $f_i$. Back to our setting, for $\alpha\subseteq\binom{[n]}{2}$, define the Fourier character
\begin{align}
\chi_{\alpha}(\s{G}) = \prod_{\{i,j\}\in\alpha}\frac{\s{G}_{ij}-q}{\sqrt{q(1-q)}},
\end{align}
for each $\s{G}\in\{0,1\}^{\binom{n}{2}}$. Then, we note that $\{\chi_{\alpha}\}_{\alpha\subseteq\binom{[n]}{2},|\alpha|\leq\s{D}}$ form an orthonormal basis for the degree-$\s{D}$ functions with respect to $\pr_{\calH_0}$. In light of \eqref{eqn:orthogonalDecomposition}, we next compute $\bE_{\calH_1}\chi_{\alpha}(\s{G})$ for each such $\alpha$. 

Fix such $\alpha$. Conditioned on the planted structure $\Gamma$, the edges of $\s{G}$ become independent, and therefore, $\bE_{\calH_1}\chi_{\alpha}(\s{G}) = \bE_{\Gamma}\prod_{\{i,j\}\in\alpha}\bE\pp{\left.\frac{\s{G}_{ij}-q}{\sqrt{q(1-q)}}\right|\Gamma}$. Let $\eta\triangleq\frac{1-q}{q}$. There are three possible cases:
\begin{itemize}
\item If $\{i,j\}\in\alpha$ is such that $i,j\not\in v(\Gamma)$, then
\begin{align}
\bE\pp{\left.\frac{\s{G}_{ij}-q}{\sqrt{q(1-q)}}\right|i,j\not\in v(\Gamma)} = 0,
\end{align}
since if $i$ or $j$ is not in $\Gamma$, then the edge $\{i,j\}$ is included in $\s{G}$ with probability $q$.
\item If $\{i,j\}\in\alpha$ is such that $i,j\in v(\Gamma)$  and $\{i,j\}\in e(\Gamma)$, then
\begin{align}
\bE\pp{\left.\frac{\s{G}_{ij}-q}{\sqrt{q(1-q)}}\right|i,j\in v(\Gamma),\{i,j\}\in e(\Gamma)} = \sqrt{\frac{1-q}{q}} = \sqrt{\eta}.
\end{align}
\item If $\{i,j\}\in\alpha$ is such that $i,j\in v(\Gamma)$  and $\{i,j\}\not\in e(\Gamma)$, then
\begin{align}
\bE\pp{\left.\frac{\s{G}_{ij}-q}{\sqrt{q(1-q)}}\right|i,j\in v(\Gamma),\{i,j\}\not\in e(\Gamma)} = -\sqrt{\frac{q}{1-q}} =-\frac{1}{\sqrt{\eta}}.
\end{align}
\end{itemize}
Let $\bar{v}(\alpha) = \cup_{\{v_1,v_2\}\in\alpha}\{v_1,v_2\}$ be the vertex set of the edges in $\alpha$. The it is clear from the above that the conditional expectation is non-zero only if $\bar{v}(\alpha)\subseteq v(\Gamma)$. Concluding,
\begin{align}
\bE\pp{\left.\frac{\s{G}_{ij}-q}{\sqrt{q(1-q)}}\right|\Gamma} =   \Ind\pp{\{i,j\}\in\alpha,v(\Gamma)}\cdot [\sqrt{\eta}]^{\Ind\pp{\{i,j\}\in e(\Gamma)}}\pp{-\frac{1}{\sqrt{\eta}}}^{\Ind\pp{\{i,j\}\not\in e(\Gamma)}}. 
\end{align}
Let $\s{Int}(\alpha,\Gamma)\triangleq\sum_{\{i,j\}\in\alpha}\Ind\pp{\{i,j\}\in e(\Gamma)}$ and $\overline{\s{Int}}(\alpha,\Gamma)\triangleq\sum_{\{i,j\}\in\alpha}\Ind\pp{\{i,j\}\not\in e(\Gamma)}$. Then,
\begin{align}
\prod_{\{i,j\}\in\alpha}\bE\pp{\left.\frac{\s{G}_{ij}-q}{\sqrt{q(1-q)}}\right|\Gamma} &= \Ind\pp{\bar{v}(\alpha)\subseteq v(\Gamma)}\eta^{\frac{1}{2}\s{Int}(\alpha,\Gamma)}\pp{-\frac{1}{\sqrt{\eta}}}^{\overline{\s{Int}}(\alpha,\Gamma)}\\
& = \Ind\pp{\bar{v}(\alpha)\subseteq v(\Gamma)}\eta^{-\frac{1}{2}|\alpha|}\pp{-\eta}^{{\s{Int}}(\alpha,\Gamma)}.
\end{align}
Next, we average over $\Gamma$. We have
\begin{align}
|\bE_{\calH_1}\chi_{\alpha}(\s{G})| &\leq \eta^{-\frac{1}{2}|\alpha|} \bE\pp{\Ind\pp{\bar{v}(\alpha)\subseteq v(\Gamma)}\eta^{\s{Int}(\alpha,\Gamma)}}.\label{eqn:PolyMethodRight}
\end{align}
While in principle one can analyze the expectation term at the r.h.s. of \eqref{eqn:PolyMethodRight}, it turns out that in the regime where $q$ is near constant, we can focus on the extreme cases where $\Gamma_k$ is either a clique or an independent set. Specifically, for $q<1/2$, we have $\eta = (1-q)/q>1$, and therefore, 
\begin{align}
\inf_{\Gamma_k}|\bE_{\calH_1}\chi_{\alpha}(\s{G})|&\leq \inf_{\Gamma_k}\eta^{-\frac{1}{2}|\alpha|} \bE\pp{\Ind\pp{\bar{v}(\alpha)\subseteq v(\Gamma)}\eta^{\s{Int}(\alpha,\Gamma)}}\\
& = \eta^{-\frac{1}{2}|\alpha|}\pr\pp{\bar{v}(\alpha)\subseteq v(\Gamma)},\label{eqn:PolyIndep1}
\end{align}
where the minimum is achieved by taking $\Gamma_k$ to be an independent set. On the other hand,
\begin{align}
\sup_{\Gamma_k}|\bE_{\calH_1}\chi_{\alpha}(\s{G})|&\leq \sup_{\Gamma_k}\eta^{-\frac{1}{2}|\alpha|} \bE\pp{\Ind\pp{\bar{v}(\alpha)\subseteq v(\Gamma)}\eta^{\s{Int}(\alpha,\Gamma)}}\\
& = \eta^{\frac{1}{2}|\alpha|}\pr\pp{\bar{v}(\alpha)\subseteq v(\Gamma)},\label{eqn:PolyClicue1}
\end{align}
and the maximum is achieved by taking $\Gamma_k$ to be a clique. For $q>1/2$ we have $\eta<1$, and accordingly,
\begin{align}
&\inf_{\Gamma_k}|\bE_{\calH_1}\chi_{\alpha}(\s{G})|\leq\eta^{\frac{1}{2}|\alpha|}\pr\pp{\bar{v}(\alpha)\subseteq v(\Gamma)},\label{eqn:PolyClicue2}
\end{align}
achieved by a clique, and
\begin{align}
&\sup_{\Gamma_k}|\bE_{\calH_1}\chi_{\alpha}(\s{G})|\leq\eta^{-\frac{1}{2}|\alpha|}\pr\pp{\bar{v}(\alpha)\subseteq v(\Gamma)},\label{eqn:PolyIndep2}
\end{align}
achieved by an independent set. In the following, we focus on the case where $q<1/2$ and analyze the r.h.s. of \eqref{eqn:PolyClicue1}, keeping in mind that the other cases can be handled in the same way. Note that the probability that $\bar{v}(\alpha)\subseteq v(\Gamma)$ is clearly $\binom{k}{|\bar{v}(\alpha)|}/\binom{n}{|\bar{v}(\alpha)|}$. Therefore, for $q<1/2$ and $\Gamma_k$ representing a clique, we have
\begin{align}
|\bE_{\calH_1}\chi_{\alpha}(\s{G})|&\leq\eta^{\frac{1}{2}|\alpha|}\frac{\binom{k}{|\bar{v}(\alpha)|}}{\binom{n}{|\bar{v}(\alpha)|}}\\
&\leq \eta^{\frac{1}{2}|\alpha|}\p{\frac{ek}{n}}^{|\bar{v}(\alpha)|},
\end{align}
where we have used the fact that $(y/x)^x\leq\binom{y}{x}\leq (ey/x)^x$. Thus, using \eqref{eqn:orthogonalDecomposition} we get 
\begin{align}
\norm{\s{L}_{n,\leq\s{D}}}_{\calH_0}^2 &\leq \sum_{0<|\alpha|\leq\s{D}}\eta^{|\alpha|}\p{\frac{ek}{n}}^{2|\bar{v}(\alpha)|}.\label{eqn:orthogonalDecomposition2}
\end{align}
Now, for any set $\calV\subseteq[n]$, we have
\begin{align}
\sum_{0<|\alpha|\leq\s{D}:\bar{v}(\alpha)=\calV}\eta^{|\alpha|}\p{\frac{ek}{n}}^{2|\bar{v}(\alpha)|}
& = \p{\frac{ek}{n}}^{2|\calV|}\cdot\sum_{0<|\alpha|\leq\s{D}:\;\bar{v}(\alpha)=\calV}\eta^{|\alpha|}\\
& \leq \p{\frac{ek}{n}}^{2|\calV|}\cdot\sum_{0<|\alpha|\leq\s{D}:\;\alpha\subseteq\binom{\calV}{2}}\eta^{|\alpha|}\\
& = \p{\frac{ek}{n}}^{2|\calV|}\cdot\sum_{\ell=1}^{\min\p{\s{D},\binom{|\calV|}{2}}}\binom{\binom{|\calV|}{2}}{\ell}\eta^{\ell}\\
&\leq \p{\frac{ek}{n}}^{2|\calV|}\cdot\sum_{\ell=1}^{\min\p{\s{D},\binom{|\calV|}{2}}}\pp{\binom{|\calV|}{2}\eta}^{\ell}\\
&\leq \p{\frac{ek}{n}}^{2|\calV|}\p{1+\binom{|\calV|}{2}\eta}^{\min\p{\s{D},\binom{|\calV|}{2}}},
\end{align}
where in the last inequality we have used the fact that $\sum_{i=0}^k\binom{n}{i}\leq\sum_{i=0}^k n^i1^{k-i}\leq(1+n)^k$. Next, every $\alpha$ with $|\alpha|\leq\s{D}$ has $|\bar{v}(\alpha)|\leq 2\s{D}$. Also, there are $\binom{n}{t}\leq n^t$ sets $\calV\subseteq[n]$ sets $\calV$ such that $|\calV|=t$. Thus,
\begin{align}
\norm{\s{L}_{n,\leq\s{D}}}_{\calH_0}^2 &\leq\sum_{t=2}^{2\s{D}}\p{\frac{ek}{n}}^{2t}n^t(1+t^2\eta/2)^{\min\p{\s{D},t^2/2}}\\
& \leq \sum_{t=2}^{2\s{D}}\p{\frac{e^2k^2}{n}}^{t}\pp{t\sqrt{\eta}}^{2\min\p{\s{D},t^2/2}}\\
&= \sum_{t\leq\sqrt{2\s{D}}}\p{\frac{e^2k^2}{n}}^{t}\pp{t\sqrt{\eta}}^{t^2}+\sum_{\sqrt{2\s{D}}<t<2\s{D}}\p{\frac{e^2k^2}{n}}^{t}\pp{t\sqrt{\eta}}^{2\s{D}}.\label{eqn:normL2upper}
\end{align}
Denote the summand in the first term at the r.h.s. of \eqref{eqn:normL2upper} by $\s{T}_t$. Then, applying the ratio test on the first term we get
\begin{align}
\frac{\s{T}_{t+1}}{\s{T}_t} \leq \frac{e^2k^2}{n} e^t(1+t)^{2t+1}\eta^{\frac{1}{2}+t}\leq \frac{e^2k^2}{n} e^{\sqrt{2\s{D}}}(1+\sqrt{2\s{D}})^{2\sqrt{2\s{D}}+1}\eta^{\frac{1}{2}+\sqrt{2\s{D}}}.\label{eqn:polyMethod1}
\end{align}
If \eqref{eqn:polyMethod1} is upper bounded by a constant strictly less than one, then the first term at the r.h.s. of \eqref{eqn:normL2upper} is upper bounded by a constant. The same is true for the second term at the r.h.s. of \eqref{eqn:normL2upper}. Therefore, for \eqref{eqn:normL2upper} to be bounded as $n\to\infty$ we need
\begin{align}
\frac{e^2k^2}{n} e^{\sqrt{2\s{D}}}(1+\sqrt{2\s{D}})^{2\sqrt{2\s{D}}+1}\eta^{\frac{1}{2}+\sqrt{2\s{D}}}<c<1,\label{eqn:condClique1}
\end{align}
for some $0<c<1$, and we used the fact that for $q<1/2$ we have $\eta>1$. We can repeat the same calculation above to the other cases in \eqref{eqn:PolyIndep1}, \eqref{eqn:PolyClicue2}, and \eqref{eqn:PolyIndep2}.  Specifically, for \eqref{eqn:PolyIndep1} we get the condition:
\begin{align}
\frac{e^2k^2}{n} e^{\sqrt{2\s{D}}}(1+\sqrt{2\s{D}})^{2\sqrt{2\s{D}}+1}\eta^{-\frac{1}{2}}<c<1,\label{eqn:condIndep1}
\end{align}
for \eqref{eqn:PolyClicue2}, we get:
\begin{align}
\frac{e^2k^2}{n} e^{\sqrt{2\s{D}}}(1+\sqrt{2\s{D}})^{2\sqrt{2\s{D}}+1}\eta^{\frac{1}{2}}<c<1,\label{eqn:condClique2}
\end{align}
and finally for \eqref{eqn:PolyIndep2}, we get:
\begin{align}
\frac{e^2k^2}{n} e^{\sqrt{2\s{D}}}(1+\sqrt{2\s{D}})^{2\sqrt{2\s{D}}+1}\eta^{-\frac{1}{2}-\sqrt{2\s{D}}}<c<1.\label{eqn:condIndep2}
\end{align}
Now, recall that for $q<1/2$ the conditions in \eqref{eqn:condClique1} and \eqref{eqn:condIndep1} represent the extreme cases of planted clique and independent set, respectively; to wit, the computational barrier of any planted structure $\Gamma_k$ lies between the computational barriers of those two structures. However, it can be seen that when $\s{D} = C\log n$, for every $C>0$, and $k=n^{1/2-\epsilon}$, for every $\epsilon>0$, both the conditions in \eqref{eqn:condClique1} and \eqref{eqn:condIndep1} hold, assuming that $\frac{1}{n^{o(1)}}\leq q\leq1-\frac{1}{n^{o(1)}}$. The same is true for $q>1/2$, which concludes the proof. 

\section{Conclusion and Outlook}\label{ref:conc_out}

This work proposes a new model for inference of general combinatorial structures planted in random graphs. For this model we provided a thorough analysis of the fundamental limits from both statistical and computational perspectives, when $q$ is near constant. There are several exciting directions for future work. Specifically, a major goal going forward is to derive the statistical and computational limits for a general scaling of $q$ with $n$. Finding the correct dependency of the statistical and computational barriers on the structure $\Gamma_k$ for a general scaling of $q$ seems challenging, and we are currently investigating this direction. Another interesting problem for future research is the case of random hidden structures, i.e., the edges $\calE_{\Gamma_k}$ of the planted structure are drawn at random with probability $p>0$. For such a model, it will be interesting to understand how the statistical and computational barriers change as a function of $p,q,k$, and the structure $\Gamma_k$. Other important generalizations are: multiple (disjoint or overlapping) hidden structures, adversarial models where an adversary is allowed to remove edges outside the planted structure before the graph is observed by the learner, and general (not necessarily binary) uniform and planting measures. Finally, note that in this paper we focused on the computational barriers of the detection problem. Nonetheless, we would like to emphasize that the total degree and spectral tests can be converted to recovery algorithms as in, e.g., \cite{chen2016statistical,alon1998finding}. For example, the structure $\Gamma_k$ can be estimated by taking the $k$ vertices whose degree (in $\s{G}$ or $\s{G}^c$) is the maximal. One can show that this recovery algorithm is successful as long as $k\geq\Omega(\sqrt{n\log n})$, while a spectral recovery algorithm as in  \cite{alon1998finding} can recover $\Gamma_k$ if $k\geq\Omega(\sqrt{n})$.



\bibliographystyle{unsrt}
\bibliography{bibfile}
\end{document}

%% file: macros.tex
\newcommand{\bE}{\mathbb{E}}

\newtheorem{definition}{Definition}

\newtheorem{theorem}{Theorem}
\newtheorem{corollary}{Corollary}
\newtheorem{conjecture}{Conjecture}
\newtheorem{lemma}{Lemma}

\newenvironment{fminipage}%
  {\begin{Sbox}\begin{minipage}}%
  {\end{minipage}\end{Sbox}\fbox{\TheSbox}}

%% file: myshorts.tex
\newcommand {\pr} {\mathbb{P}}

\newcommand{\calE}{{\cal E}}

\newcommand{\calG}{{\cal G}}
\newcommand{\calH}{{\cal H}}

\newcommand{\calK}{{\cal K}}
\newcommand{\calL}{{\cal L}}

\newcommand{\calN}{{\cal N}}

\newcommand{\calP}{{\cal P}}

\newcommand{\calS}{{\cal S}}

\newcommand{\calV}{{\cal V}}

\newcommand{\be}{\begin{equation}}
\newcommand{\ee}{\end{equation}}
\newcommand{\beqna}{\begin{eqnarray}}
\newcommand{\eeqna}{\end{eqnarray}}
